\documentclass[12pt]{iopart}
\usepackage{graphicx}

\begin{document}

\title{Two-proton radioactivity}

\author{Bertram Blank \ddag ~ and Marek P{\l}oszajczak \dag}

\address{\ddag~ Centre d'Etudes Nucl\'eaires de Bordeaux-Gradignan - Universit\'e Bordeaux I - CNRS/IN2P3, 
         Chemin du Solarium, B.P. 120, 33175 Gradignan Cedex, France\\
         \dag~ Grand Acc\'el\'erateur National d'Ions Lourds (GANIL),
         CEA/DSM-CNRS/IN2P3, BP 55027, 14076 Caen Cedex 05, France}

\begin{abstract}

In the first part of this review, experimental results  which lead to the 
discovery of two-proton radioactivity are examined.
Beyond two-proton emission from nuclear ground states, we also discuss experimental studies
of two-proton emission from excited states populated either by
nuclear $\beta$ decay or by inelastic reactions. In the second part, we review the modern 
theory of two-proton radioactivity.  An outlook to future experimental studies and theoretical 
developments will conclude this review.

\end{abstract}


\submitto{\RPP}

\vspace*{1.0cm}
\begin{center}
Version: \today
\end{center}
\maketitle

\section{Introduction}

Atomic nuclei are made of two distinct particles, the protons and the neutrons.
These nucleons constitute more than 99.95\% of the mass of an atom. In order to
form a stable atomic nucleus, a subtle equilibrium between the number of protons and
neutrons has to be respected. This condition is fulfilled for 259 different combinations
of protons and neutrons. These nuclei can be found on Earth. In addition, 26 nuclei form
a quasi stable configuration, i.e. they decay with a half-life comparable or longer
than the age of the Earth and are therefore still present on Earth.

These stable or quasi-stable nuclei are characterised by an approximately equal number of neutrons
and protons for light nuclei, whereas going to heavier nuclei the neutron excess increases to reach 
N-Z=44 for the heaviest stable nucleus $^{208}$Pb. 
In addition to these 285 stable or quasi-stable nuclei, some 4000-6000
unstable nuclei are predicted to exist, the exact number depending on the theoretical model used.
Close to 2500 nuclei have been observed already. These unstable
radioactive nuclei can be classified into seven categories depending on their decay modes:
i) $\alpha$ emitters (about 375) which decay by ejecting a $^4$He nucleus, ii) $\beta^+$ emitters (1040)
which transform a proton into a neutron plus a positron and a neutrino, or by capturing 
an electron from their atomic shells and emitting a neutrino, iii) $\beta^-$ emitters (1020) which
transform a neutron into a proton by emitting an electron and an anti-neutrino, iv) fissioning
nuclei (30) which split up into two roughly equal pieces, v) one-proton (1p) emitters (about 25), vi) 
two-proton (2p) emitters (2) and, finally, (vii) exotic-cluster emitters (about 25 different 
nuclei emitting fragments of C, O,  F, Ne, Mg, and Si). 
These  exotic radioactivities with masses from 14 to 32 may be considered as intermediate between $\alpha$ decay and fission.
However, it should be mentioned that this decay mode is never dominant.

Although some radioactive nuclei can decay by two or three of these different decay modes, 
the main decay mode is dictated by the proton-to-neutron ratio. For example, 
nuclei with a modest proton excess decay by $\beta^+$ disintegration. If the proton excess
increases, the nuclear forces can no longer bind all protons and one reaches the proton drip line 
where the spontaneous emission of protons takes place from the ground state.
Heavier neutron-deficient isotopes will preferentially decay by $\alpha$ emission, and for the 
heaviest nuclei fission is the dominant decay mode. For some of the heaviest nuclei, 
exotic-cluster emission is a small decay branch.

The study of these different decay types started with the discovery of radioactivity
by H. Becquerel in 1896. Following the work of Pierre and Marie Curie, E.~Rutherford in 1899 
classified the observed radioactivity into two different decay modes ($\alpha$ 
and $\beta$ rays) according to the detected particle in the decay. P.~Villard discovered 
$\gamma$ radiation in 1900 as a third radioactive decay mode. Finally, the nuclear fission 
process was observed for the first time by O.~Hahn and F.~Strassmann and correctly explained by L.~Meitner in 1938.

These "classical" decay modes have helped largely to understand the structure of
the atomic nucleus and the forces acting in the nucleus. They have also found many applications
which range from solid-state physics to astrophysics and medicine. With the discovery of a large number
of new radioactive isotopes the understanding of nuclear structure advanced as well. 

Beginning of the 1960's, Goldanskii~\cite{goldanskii60}, Zel'dovich~\cite{zeldovich60}, 
and Karnaukhov~\cite{karnaukhov61} proposed new types of proton radioactivity in very proton-rich nuclei. 
For nuclei with an odd number of protons, one-proton
radioactivity was predicted. This decay mode was indeed observed at the beginning of the 1980's in experiments
at GSI, Darmstadt~\cite{hofmann82,klepper82}. Today about 25 ground-state one-proton emitters are known~\cite{blank07review}.
Proton emission was also observed in many cases from long-lived excited states, 
the first being $^{53m}$Co~\cite{jackson70,cerny70,cerny72}.
The study of one-proton radioactivity allowed testing the nuclear mass surface, to determine the sequence of 
single-particle levels and the detailed structure of the wave function of the emitted proton ("$j$ content"
of the wave function), and to investigate nuclear deformation beyond the proton drip line.

Zel'dovich~\cite{zeldovich60} was probably the first to mention the possibility that nuclei may emit a pair of protons. 
However, it was Goldanskii~\cite{goldanskii60,goldanskii61,goldanskii62,goldanskii65,goldanskii66} and 
J\"anecke~\cite{jaenecke65} who first tried to determine candidates for 2p radioactivity. 
Many others followed and the latest predictions~\cite{brown91,cole96,ormand96} determined that $^{39}$Ti, 
$^{42}$Cr, $^{45}$Fe, and $^{49,48}$Ni should be the best candidates to discover this new decay mode.
Goldanskii~\cite{goldanskii60} also coined the name "two-proton radioactivity" and it was Galitsky and 
Cheltsov~\cite{galitsky64} who proposed a first theoretical attempt to describe the process of 2p emission. 
More than 40 years after its theoretical proposal, 2p radioactivity was 
discovered~\cite{giovinazzo02,pfuetzner02} in the decay of $^{45}$Fe. In later experiments, $^{54}$Zn~\cite{blank05zn54} and 
most likely $^{48}$Ni~\cite{dossat05} were also shown to decay by two-proton radioactivity. The observation of two-proton
emission from a long-lived nuclear ground state was preceded by the observation of two-proton emission from 
very short-lived nuclear ground state, e.g. in the cases of $^{6}$Be~\cite{bochkarev89} and $^{12}$O~\cite{kryger95}, 
and by the emission of two protons from excited states populated by $\beta$ decay~\cite{cable83} or inelastic 
reactions~\cite{bain96}. A particular case is the emission of two protons from an isomer in $^{94}$Ag$^m$~\cite{mukha06}.

The 2p emission process leads from an initial unbound state to a final state of separated fragments 
interacting by the Coulomb force. This transition to the final state can be realized in 
many ways, such as the sequential two-body decay via an intermediate resonance, the virtual 
sequential two-body decay via the correlated continuum of an intermediate nucleus, the direct 
decay into the three-particle continuum or its particular diproton ('$^2$He') limit of two sequential 
two-body emissions. 

Different types of theoretical approaches exist which focus on the three-body decay width: 
the extended R-matrix theory~\cite{brown02,brown03}, the real-energy continuum shell 
model (the so-called Shell Model Embedded in the Continuum (SMEC) 
~\cite{bennaceur99,bennaceur00,okolowicz03,rotureau05,rotureau06}) 
describing sequential two-body emission, virtual sequential two-body emission and diproton emission via two sequential 
two-body emissions~\cite{rotureau05,rotureau06}, the Gamow Shell Model (GSM) with no separation between two- and 
three-body decays from the multi-particle continuum~\cite{michel04}, the three-body models with outgoing flux 
describing direct decay into the continuum~\cite{grigorenko03f}, the three-body models combined with complex 
coordinate scaling~\cite{kurokawa04}, or the Faddeev equations combined with either outgoing flux or complex 
scaling~\cite{garrido04}. 

In the 2p decay, the three-body decay width provides global information with weighted contributions of different 
decay paths leading from the initial to the final state. Hence, the value of the width by itself does not reveal 
the dominant transition mechanism. In the theoretical analysis however, one can switch off certain decay paths and 
look for the three-body partial decay width resulting from a particular decay path. This strategy is used in the 
SMEC analysis of the experimental results. Obviously,  the conclusions of such partial studies of decay mechanisms 
are not unambiguous.

In the present review, we will summarize the state of the art of experimental and theoretical studies on two-proton
radioactivity. Previous reviews, written before the actual discovery of ground-state two-proton radioactivity, can be found 
in~\cite{goldanskii66a,aysto89}.

\section{Basic concepts for one- and two-proton radioactivity}

\begin{figure}[hht]
\begin{center}
\includegraphics[scale=0.4,angle=0]{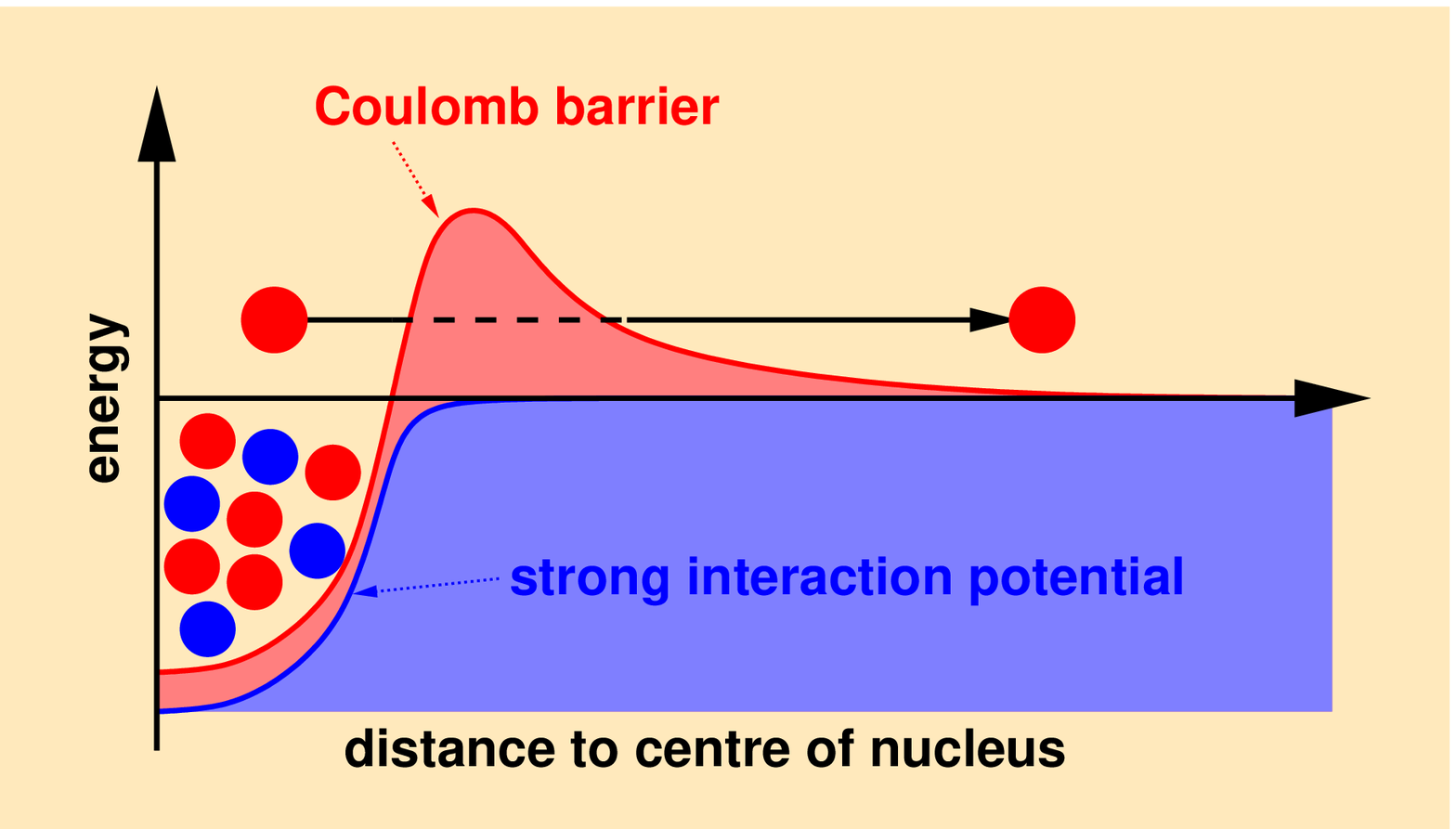} \ \hfill \
\includegraphics[scale=0.4,angle=0]{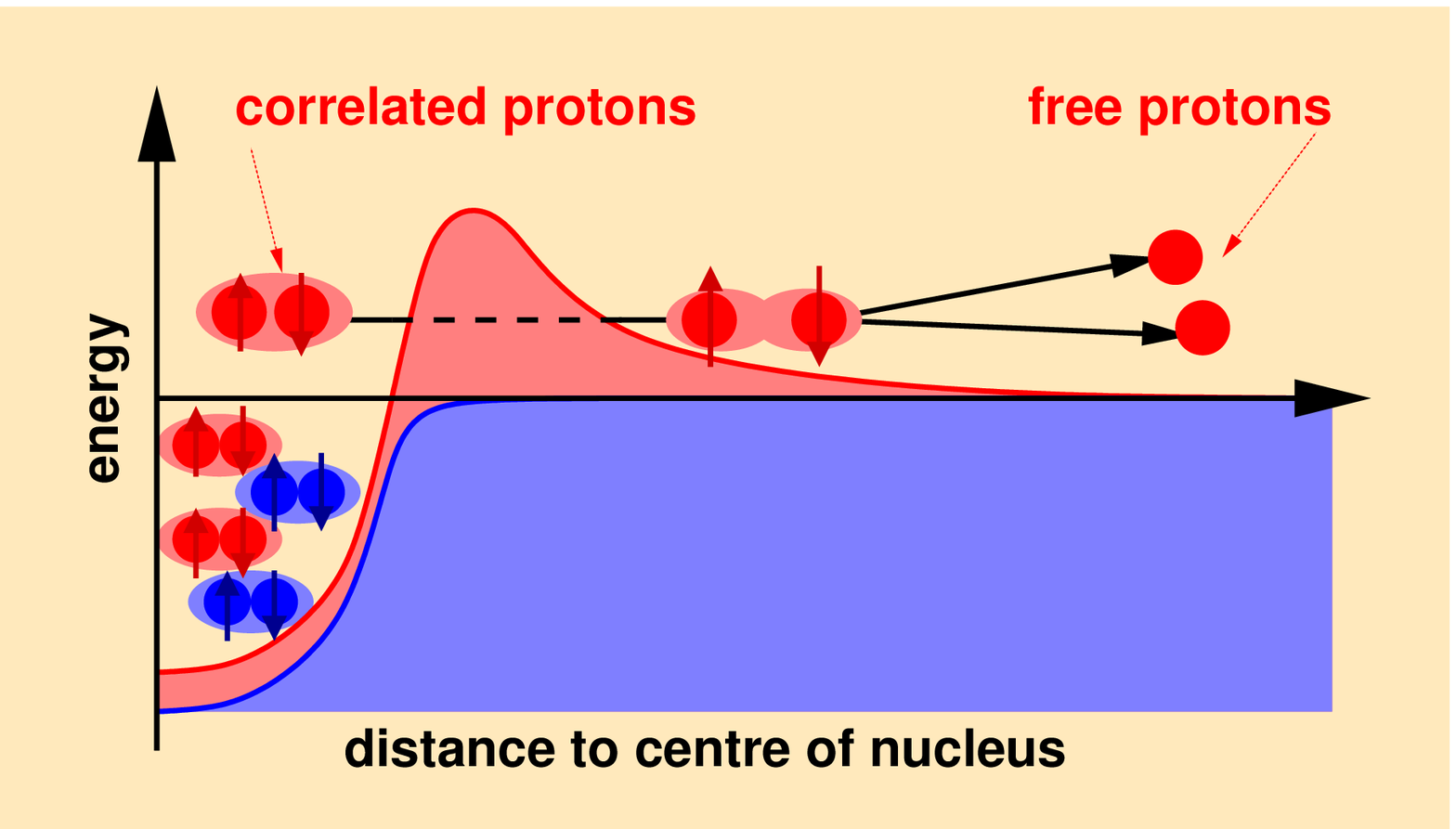}
\end{center}
\caption{Due to their charge the protons have to tunnel through the Coulomb barrier 
         generated by other protons. For one-proton emission (left),
         the tunnelling depends mainly on the barrier height. For two-proton
         emission (right) the correlation between the two protons most likely influences
         the tunnelling process. Figure courtesy of J. Giovinazzo.}
\label{fig:barrier}
\end{figure}

\begin{figure}[hht]
\begin{center}
\includegraphics[scale=0.5,angle=-90]{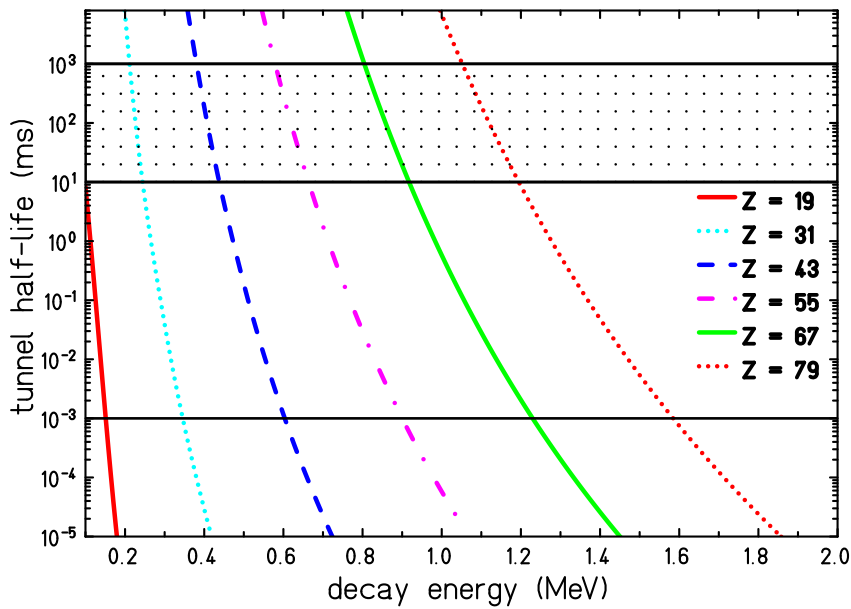}
\end{center}
\caption{Barrier penetration half-lives for a proton as a function of the nuclear charge
         and the decay energy. The half-lives are calculated from Coulomb 
         wave functions using the Wigner single-particle width (see 
         e.g.~\cite{brown91}). The horizontal line gives the lower detection time limit, 
         whereas the hatched area gives typical $\beta$-decay half-lives.}
\label{fig:tunnel}
\end{figure}

Protons are charged particles, therefore they are sensitive to the charge of other protons
which create a Coulomb barrier. This barrier prevents protons from quickly leaving the 
atomic nucleus even if they are unbound. The tunnelling probability  depends on the available energy and 
the height of the Coulomb barrier, which in turn depends on the nuclear charge Z (number of protons) 
(see Fig.~\ref{fig:barrier}). The barrier penetration
can give rise to measurable half-lives, if a certain balance between the available decay energy
and the barrier height is respected. Fig.~\ref{fig:tunnel} shows in a simple model the relation
between barrier penetration half-life and decay energy for different nuclear
charges~Z. In general, the higher is the available energy, the shorter is the tunnelling time. In turn, 
for higher Z  more energy is needed for the same tunnelling time.

The delay associated with the  tunnelling process allows for the observation of 
1p  and 2p radioactivity. Even if protons are unbound by e.g. 1~MeV, the  
tunnelling of the combined Coulomb and centrifugal barriers is not instantaneous, i.e. 
the nuclear decay is delayed by a measurable amount of time. However, due to experimental
constraints, mainly linked to the techniques used to study 1p or 2p radioactivity
and due to the competition with $\beta^+$ decay, observation limits exist. The lower half-life
limit of about 1~$\mu$s comes from the fact that often the observation of the 1p or
2p emitter is accomplished with the same detection setup which is used to detect the decay
of these nuclei. This means that a decay signal of typically 1~MeV has to be observed 
a very short time after an identification signal (i.e. an implantation signal in a silicon
detector) of several hundred MeV. This observation limit has been reached 
in 1p-emitter studies (see Sect. 3). The upper limit (see Fig.~\ref{fig:tunnel}) 
strongly depends on the structure of
the decaying nucleus which governs the $\beta$-decay half-life. It can range from a few 
milliseconds to a few seconds. To observe charged-particle emission, the barrier-penetration
half-life should be comparable or shorter than the $\beta$-decay half-life. 

In the 1p decay, there are only two particles in the final state and the decay is a 
simple back-to-back decay, where the energy is shared between the two partners, the heavy
recoil and the emitted proton, according to energy and momentum conservation. 

In the case of 2p decay, the situation is more complicated (see Fig.~\ref{fig:barrier}). 
The decay characteristics depend sensitively on the decay pattern itself. Two schematical pictures 
are usually used to represent possible limiting cases: i) the three-body decay and ii) the diproton decay. 
In the former case, the two protons do not have any correlation beyond the phase-space 
constraints, which means that only energy and momentum conservation have to be respected.
Such a decay pattern yields an isotropic angular distribution of protons which share
the total decay energy, with individual proton energies ranging from zero to the total decay energy. 
A more realistic discussion of the tunnelling process changes this picture because 
the barrier penetration strongly favours an emission of  two protons with similar energies.
In the latter case of diproton emission, one assumes that a pre-formed '$^2$He'-cluster
penetrates the Coulomb barrier and decays outside the barrier. The decay half-life depends sensibly 
on the $^2$He resonance energy. The diproton decay 
corresponds to two sequential binary decays for which the kinematics is rather easy.

These two limits of 2p decay are not very realistic, but they are easy to grasp
and give at least a schematic idea about the 2p emission. 
A theoretical description of 2p radioactivity (see Sect. 8)
is more sophisticated, although none of the models provide an unconstrained description of the nuclear 
structure and the decay dynamics involved in the 2p decay.

Up to now, only total decay energies and half-lives have been
measured in 2p radioactivity experiments. A deeper insight into the decay mechanism can be provided 
by the measurements of individual proton energies and the proton-proton angle in the center-of-mass. 
To access these observables, specific detectors have
to be developed (see Sect. 6).

\section{The discovery of one-proton radioactivity}

Soon after the prediction of proton emission from atomic nuclei, the search 
for these new types of radioactivity started. The first exotic decay
mode discovered was the $\beta$-delayed emission of one proton in the decay of $^{25}$Si~\cite{barton63}.
A similar type of proton emission from an excited state was observed 
in the search for $\beta$-delayed proton emission from $^{53}$Ni.
Instead of populating $^{53}$Ni, the authors~\cite{jackson70,cerny70}
observed the population and decay of $^{53}$Co$^m$. This isomer (T$_{1/2}$~= 247~ms) decays 
with a branching ratio of 1.5\% by emission of a 1.59(3)~MeV proton.

The first case of ground-state one-proton radioactivity was reported by Hofmann 
et al.~\cite{hofmann82} and by Klepper et al.~\cite{klepper82}.
Both experiments were conducted at GSI, the first one
at the velocity filter SHIP using a fusion-evaporation reaction with a $^{58}$Ni 
beam and a $^{96}$Ru target to produce the proton emitter $^{151}$Lu, 
the second one at the on-line mass separator with the reaction  $^{58}$Ni + $^{92}$Mo
to synthesize $^{147}$Tm.

One-proton radioactivity studies have developed soon into a powerful tool to
investigate nuclear structure close to and beyond the proton drip 
line~\cite{woods97,rykaczewski02,blank07review} with now more than thirty different nuclei
known to emit protons from their ground states or low-lying isomeric states.
These studies have given valuable information on the  sequence of single-particle levels and
their energies in the vicinity of the proton drip-line, on the $j$ content of the
nuclear wave function, on the nuclear deformation, and they allow to test
predictions of nuclear mass models in this region. Most of this information
can be obtained only by means of 1p radioactivity studies.

More recently proton radioactivity was also used to identify nuclei and to tag
events. In experiments e.g. at the Fragment Mass Separator of the Argonne National Laboratory, the
observation of a decay proton is used as a trigger to observe $\gamma$ rays
emitted in the formation of the proton emitter. Therefore, not only the
decay of these exotic nuclei can be studied, but also their nuclear structure
in terms of high-spin states and their decay via yrast cascades~\cite{seweryniak01}.

The relatively large number of known ground-state proton emitters (for almost
all odd-Z elements between Z=50 and Z=83, at least one proton-radioactive 
nucleus is experimentally observed) is due to the fact that the 1p
drip line, i.e. the limit where odd-Z nuclei can no longer bind all protons, is 
much closer to the valley of stability than the two-proton drip line. As an example, we mention
that at Z=50 the T$_z = -1$ isotope $^{98}$Sn is most likely the first 
2p unbound nucleus. For Z=51, $^{103}$Sb with T$_z$=1/2 is probably proton unbound. 
This means that in this case the 2p emitter is three mass units
more exotic than the 1p emitter, which leads to orders of magnitude
lower production cross sections.

\section{Two-proton emission from very short-lived nuclear ground states}

The first experimental investigation of a possible ground-state 2p emitter
was performed by Karnaukhov and Lu Hsi-T'ink~\cite{karnaukhov65}. These authors
tried to produce $^{16}$Ne by bombarding a nickel target with a 150~MeV $^{20}$Ne
beam from the JINR, Dubna 300~cm cyclotron. The non-observation of any 2p
event led the authors to conclude that either the production cross section
is much smaller than assumed (smaller than 10$^{-5}$~b) or that 
the half-life of $^{16}$Ne is shorter than 10$^{-8}$s. 
As early as in 1963, J\"anecke predicted its half-life to be of the order
of 10$^{-19}$s~\cite{jaenecke63} and recently its half-life was given as 
9$\times$10$^{-21}$s~\cite{audi03}.

\subsection{The decay of $^6{\rm Be}$}

If we omit early measurements which used $\pi$-induced reactions in
bubble or spark chambers~\cite{charpak65,burman68}, and mass excess 
measurements performed by Ajzenberg-Selove et al.~\cite{ajzenberg-selove59},
the first successful study of a radioactive decay with an emission of two protons
was performed in experiments with $^{6}$Be~\cite{geesaman77}. In these studies, 
$^{6}$Be was produced in a $^6$Li($^3$He,t)$^6$Be reaction. The triton as
well as the $\alpha$ particle and the protons from the decay of $^{6}$Be were observed.
The ground state of $^6$Be is known to have a life-time of the order of 
10$^{-21}$s. Therefore, this nucleus can only be identified via its decay
products. The authors observed an enhancement of low-energy $\alpha$ particles
compared to a simple phase-space distribution which could not be explained 
by means of final-state interactions. However, no conclusions on the 
decay mechanism could be drawn from these observations.

A much more detailed study of the  $^6$Be decay was performed over 
several years at the Kurchatov Institute, Moscow (see e.g.~\cite{bochkarev89,bochkarev92}).
These authors measured energy and angular distributions of the emitted particles
and interpreted their data using Jacobi coordinates. The most
important conclusion of this analysis is that the
spectra can be interpreted as a "democratic" three-body decay, i.e. the decay where none 
of the two-body subsystems of this three-body final state has a width which is small in comparison with the 
transition energy between the initial nucleus and the subsystem. 

\subsection{The decay of $^{12}{\rm O}$}

The next step in the investigation of short-lived 2p ground-state emitters
was the study of $^{12}$O at the 
National Superconducting Cyclotron Laboratory (NSCL) 
of Michigan State University, East Lansing, USA~\cite{kryger95}.
In this experiment, $^{12}$O was produced by one-neutron stripping from
a $^{13}$O secondary beam. The complete reaction kinematics was reconstructed by
measuring the momentum vectors of two protons in coincidence with the momentum
of the heavy $^{10}$C residue. From these measurements, the authors could produce
an excitation energy spectrum of $^{12}$O, determine the proton-proton energy
difference and the proton-proton angle in the $^{12}$O center of mass. 

\begin{figure}[hht]
\begin{minipage}[ht]{7cm}
\resizebox{.95\textwidth}{!}{\includegraphics{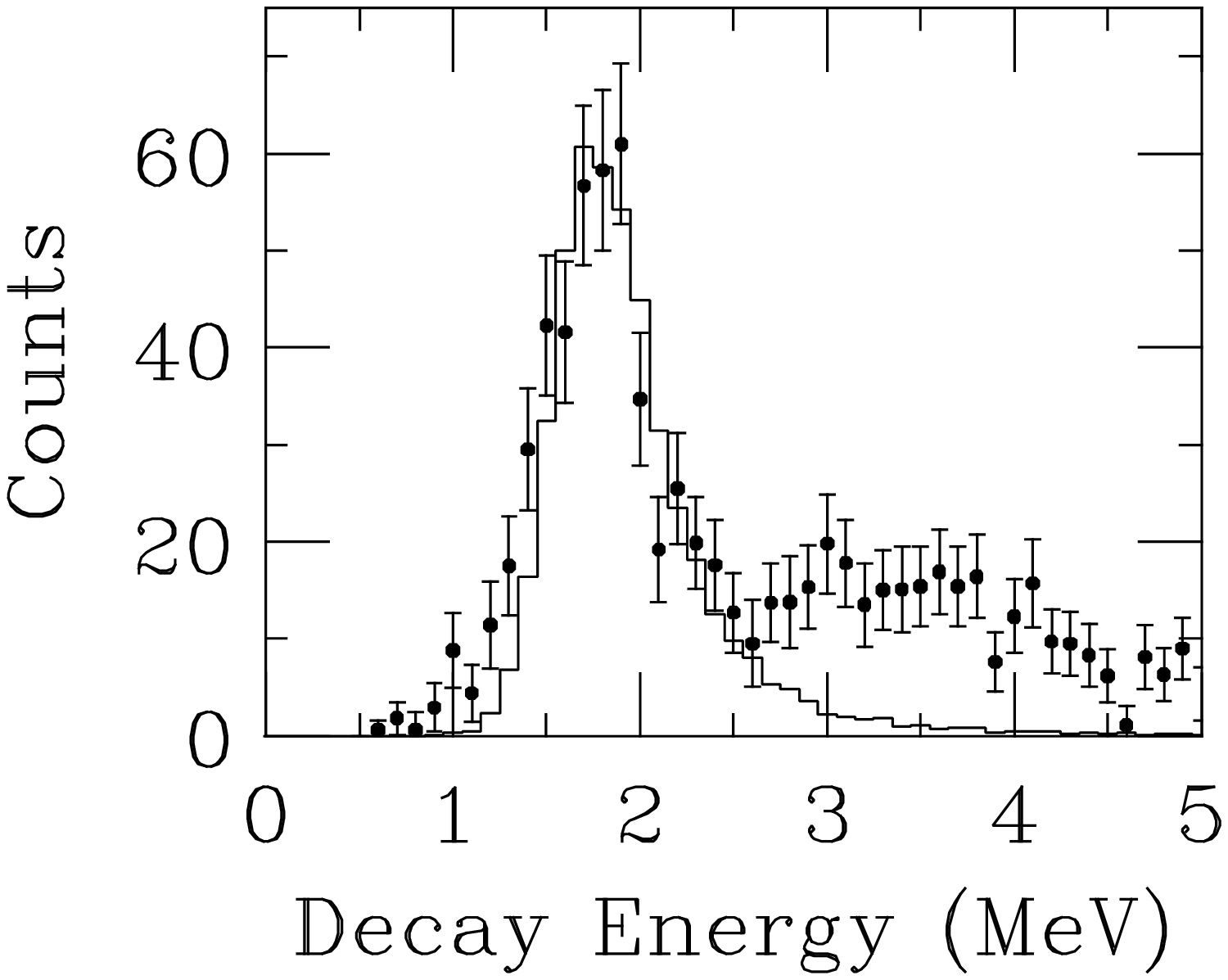}} 
\end{minipage} \ \hfill \
\begin{minipage}[ht]{7cm}
\resizebox{.95\textwidth}{!}{\includegraphics{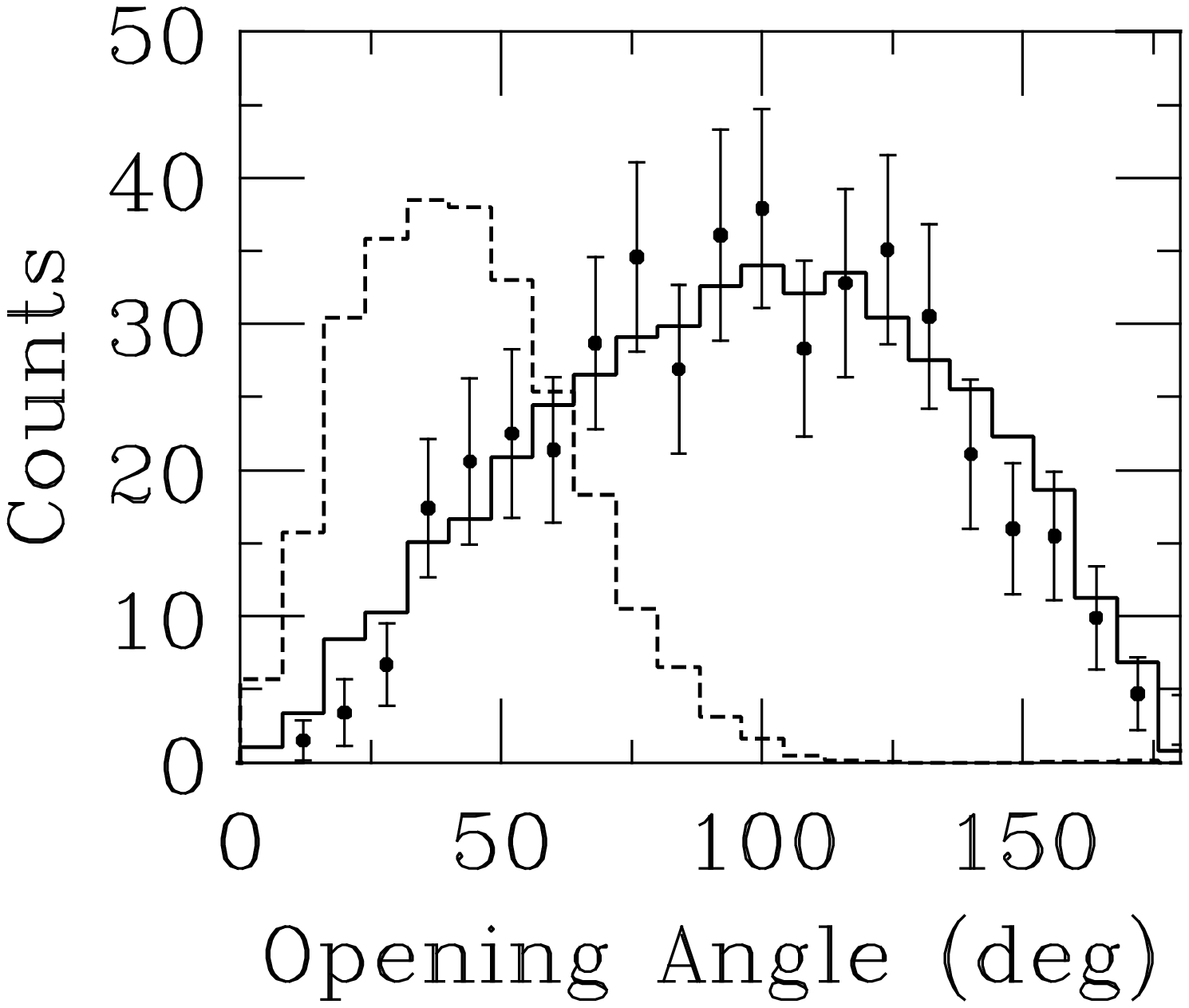}} \ \hfill
\end{minipage} \ \hfill \ 
\begin{minipage}[ht]{7cm}
\resizebox{1.\textwidth}{!}{\includegraphics[angle=-90]{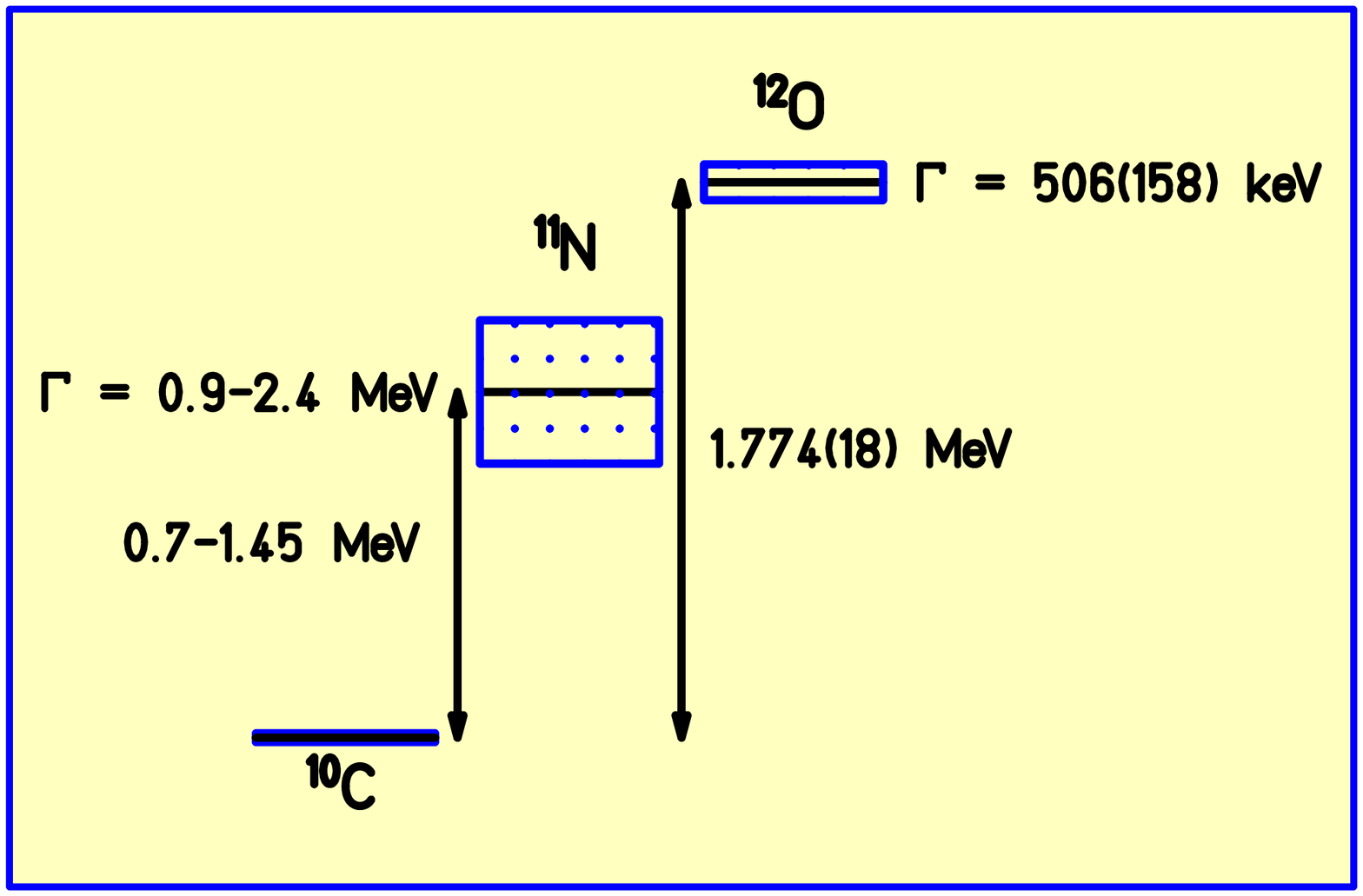}} 
\end{minipage}
\caption{Experimental spectra for the decay of $^{12}$O and level scheme. The
         experimental data are compared with Monte-Carlo simulations using the two
         extreme pictures, i.e. sequential (full line) or diproton (dashed line) decay. Quantitative agreement is obtained
         within the sequential decay picture. The decay scheme evidences that the
         sequential path is energetically open~\cite{kryger95,azhari98}. Figures re-used with permission from APS.}
\label{fig:o12}
\end{figure}

These experimental data were interpreted by means of Monte-Carlo simulations, which
employed the R-matrix theory~\cite{lane58} to describe the 2p decay in two
extreme pictures: diproton emission with a $^{2}$He resonance energy of 50~keV and
three-body decay which is only restricted by the available phase space. Fig.~\ref{fig:o12}
summarizes these results. A comparison of the experimental results and
the simulations clearly favours an uncorrelated three-body emission or a sequential
decay, which may have roughly the same characteristics. At the time
of this work, these conclusions were somewhat surprising, as it was believed
that the intermediate 1p daughter state in $^{11}$N was inaccessible due 
to energy conservation. In addition, a very large $^{12}$O decay width 
was needed to explain the experimental data. However, later 
measurements~\cite{axelsson96,lepine98,azhari98a} revealed that the $^{11}$N 
ground state lies lower in mass and opens therefore the way for a sequential 
decay. With these new results, a consistent description of the decay of 
$^{12}$O could be provided~\cite{azhari98}.

\subsection{The decay of $^{16}{\rm Ne}$ and $^{19}{\rm Mg}$}

The decay of $^{19}$Mg was recently studied in an experiment at the fragment separator FRS of 
GSI~\cite{mukha07}. In this experiment, $^{19}$Mg was produced by a one-neutron stripping reaction
at the FRS intermediate focal plane of $^{20}$Mg, produced itself at the FRS target in a fragmentation
reaction of the $^{24}$Mg primary beam. The protons emitted in the decay of $^{19}$Mg were
detected and tracked by a set of three DSSSD, whereas the heavy decay product, $^{17}$Ne, was identified  
and its momentum measured in the second half of the FRS. 

The measured data were compared to Monte Carlo simulations and allowed to extract the half-life of 
$^{19}$Mg of $T_{1/2}$~= 4.0(15)~ps and the associated two-proton Q-value:  
$Q_{2p}$~= 0.75(5)~MeV.
This value is much smaller than standard predictions like the one of the Atomic Mass Evaluation~\cite{audi03} ($Q_{2p}$~= 2.00(25)~MeV), 
the Garvey-Kelson relation~\cite{masson88} ($Q_{2p}$~= 1.02(2)~MeV), or the prescription from Antony et al.~\cite{antony97}
($Q_{2p}$~= 1.19(3)~MeV). However, the correlation between the Q value and the decay half-life is well reproduced within the three-body
model of Grigorenko and co-workers~\cite{grigorenko03f}.

In the same experiment~\cite{mukha07}, the decay of $^{16}$Ne by two-proton emission was also observed and the known decay 
Q value~\cite{woodward83} of 1.41(2)~MeV was used to test the experimental procedure.

\subsection{Conclusions and future studies with short-lived ground-state two-proton emitters}

For the systems $^{6}$Be, $^{12}$O, and $^{16}$Ne, the decaying states have large widths so that the 
2p emitter state and the 1p daughter state overlap. Therefore, the sequential decay channel is opened 
and the decay will most likely take place sequentially. The situation in the case of $^{19}$Mg is less 
clear, as the structure and the mass of the one-proton daughter $^{18}$Na is not known. 
However, with increasing nuclear charge, the Coulomb barrier becomes stronger and the wave 
function is more and more confined in the nuclear interior which yields narrower states. 

Therefore, it might well be that in nuclei like $^{21}$Si, $^{26}$S, or $^{30}$Ar the Coulomb barrier is
already strong enough so that, with a possibly small decay energy, narrow levels avoid overlapping
and the sequential decay path is, at least in a simple picture, not open.

\section{Observation of two-proton emission from excited states}

Generally speaking, the 2p emission can take place as a sequential-in-time process or a 
simultaneous process. In its standard form, the sequential process 
happens when levels in the 1p daughter nucleus, both resonant and non-resonant ones, are 
accessible for 1p emission. This situation may be encountered when the emission takes place from
highly excited levels of the 2p emitter. These levels can be fed either by
$\beta$ decay or by a nuclear reactions.

A clear signal for a sequential emission can be obtained when the intermediate
level through which the emission passes is identified. This is the case, for example, when
well defined one-proton groups sum up to a fixed 2p energy. The
one-proton groups give then the energy difference between two well-defined
nuclear levels.

\subsection{$\beta$-delayed two-proton emission}

The possibility of $\beta$-delayed 2p ($\beta$2p) emission was first
discussed by J\"anecke~\cite{jaenecke65}. He concluded that
there should be no "$\beta$-delayed diproton emitters" with Z$\leq$14.
This decay mode was also studied by Goldanskii~\cite{goldanskii80} who proposed possible 
candidates.

The first experimental observation of $\beta$-delayed 2p emission
was reported by Cable et al. in 1983~\cite{cable83}. Their experiments,
using a $^3$He induced reaction on a magnesium target and a helium-jet technique,
led to the first observation of $^{22}$Al by means of its $\beta$-delayed
1p emission from the isobaric analogue state (IAS) in $^{22}$Mg to the
ground and first excited states of $^{21}$Na~\cite{cable82}. The $\beta$2p
decay mode was subsequently observed with two silicon detector telescopes which allowed
to measure the energy of the two individual protons with good resolution~\cite{cable83}.
 
 From the observation of well defined one-proton groups, Cable et al. concluded that this decay 
is a sequential mode via intermediate states in the 1p daughter nucleus
$^{21}$Na. This conclusion was confirmed in a further experimental 
study~\cite{jahn85} where it turned out that the angular distribution of the
two protons is compatible with an isotropic emission, although a small 
angular correlation (less than 15\%) could not be excluded.

Meanwhile more $\beta$2p emitters have been identified. Shortly after
the discovery of $\beta$2p decay from  $^{22}$Al~\cite{cable83}, 
the same authors observed $\beta$2p emission
from $^{26}$P~\cite{honkanen83} and from $^{35}$Ca~\cite{aysto85}. All in all, 
nine $\beta$2p emitters have been observed: $^{22}$Al~\cite{cable83},
$^{23}$Si~\cite{blank97si23}, $^{26}$P~\cite{honkanen83}, $^{27}$S~\cite{borrel87}, 
$^{31}$Ar~\cite{reiff89}, $^{35}$Ca~\cite{aysto85}, $^{39}$Ti~\cite{moltz92}, 
$^{43}$Cr~\cite{borrel92,giovinazzo07}, and $^{50}$Ni~\cite{dossat07}. 
Most of these studies were characterized
by rather low statistics inherent to most investigations of very exotic nuclei
and hence no detailed search for angular correlations could be performed. In addition, 
most of the observed decays proceed via the IAS in the $\beta$-decay
daughter nucleus. The 2p emission from this state is forbidden by  the isospin conservation
law and can only take place via a small isospin impurity of the IAS.

The first high-statistics study of a $\beta$2p emitter was carried out at 
ISOLDE in CERN~\cite{axelsson98,axelsson98a,thaysen99,fynbo99}. For
the first time, a rather high-efficiency, high-granularity setup which
allows a high-resolution measurement in energy and angle was used. This experiment,
which studied the decay of $^{31}$Ar, demonstrated the occurrence of $\beta$2p 
decay branches via states other than the IAS. 
Indeed, many Gamow-Teller fed states could be shown to decay by 2p emission.
The main interest of these decays is that they are isospin allowed and cover
a wider range of spins for the 2p emitter states, the possible intermediate 
states and the final states.

Detailed studies of the proton-proton angular correlation for different 2p
emission branches did not yield evidence for any angular correlation. 
In fact, the results were compatible with sequential emission via intermediate
states in the 1p daughter nucleus. Although there was some activity detected
with the two protons sharing the available decay energy equally, the statistics
of these events was too low and the angular coverage of the setup was not
uniform enough to draw any conclusion on the possible observation of events,
where the protons were correlated in energy and angle. Therefore, despite
a factor of 10-100 higher statistics as compared to earlier experiments, 
no angular and energy correlation, a possible signature of a strong proton-proton
correlation, could be evidenced.

This observation is not really astonishing. According to Brown~\cite{brown90}, who studied
the competition between 1p emission and correlated 2p (diproton) emission from the
IAS  in $^{22}$Mg, the weight of the direct 2p branch is
expected to be at best of the order of a few percent and therefore rather
difficult to observe experimentally.
New attempts to search for such a correlated 2p emission in $^{31}$Ar are on the
way~\cite{adimi07}. These studies take profit from a much more uniform
angular coverage.

\subsection{Two-proton emission from excited states populated in nuclear reactions}

Besides populating excited 2p emitter states by $\beta$ decay, these nuclear 
states can also be fed by nuclear reactions like pick-up, transfer, or fragmentation.
In this section, we will describe experiments which used this kind of interaction to
study 2p emission from excited states.

\subsubsection{Two-proton emission from $^{14}${\rm O}}

\begin{figure}[hht]
\begin{center}
\includegraphics[scale=0.5,angle=-90]{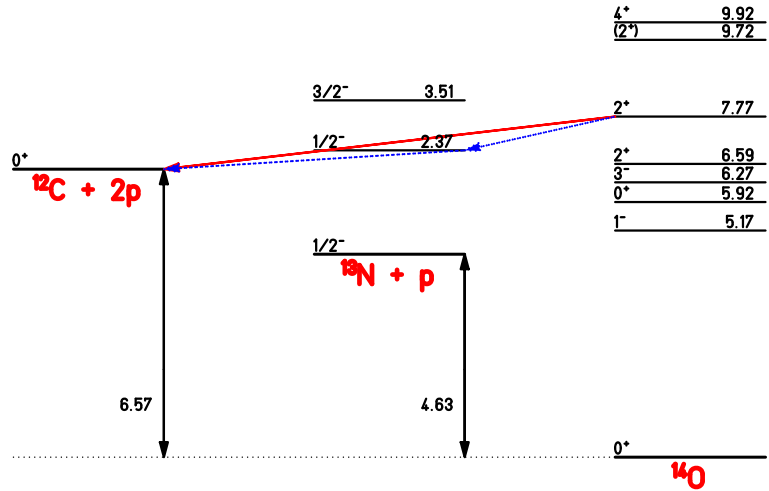} 
\end{center}
\caption{Decay scheme of the 7.77~MeV resonance in $^{14}$O by two-proton emission.
         A direct decay to the ground state of $^{12}$C is possible, but the
         decay was found to be predominantly sequential via an intermediate state at  2.37~MeV   
         in $^{13}$N.}
\label{fig:bain}
\end{figure}

In an experiment performed at Louvain-La-Neuve, Bain et al.~\cite{bain96}
bombarded a (CH$_2$)$_n$ target with a $^{13}$N radioactive beam populating 
highly-excited states in $^{14}$O. The 7.77~MeV state  in $^{14}$O, which  is 
strongly populated in 2p transfer reactions~\cite{greenfield72}, is expected to 
have a significant admixture of a 2p configuration outside of the  $^{12}$C core. 
In addition, this state is 2p unbound by 
about 1.2~MeV and, unlike states populated by a super-allowed $\beta$ decay,
the 2p decay to the ground state of $^{12}$C is isospin allowed. Therefore, 
this resonance was believed to have a small 2p decay branch, either directly to
the $^{12}$C ground state or sequentially via a narrow state in $^{13}$N
(see Fig.~\ref{fig:bain}).

The two protons from the decay of the 7.77~MeV resonance were detected with 
the LEDA device~\cite{bain96}. The data showed clearly a resonant structure when the beam 
energy was 'on-resonance'. The authors determined a 0.16(3)\% 2p branch for the decay of this resonance.
The individual energies of the two protons could be explained by a sequential 
decay pattern via the 2.37~MeV state in $^{13}$N. Monte-Carlo simulations employing
R-matrix theory showed that the data could be best explained with a 100\% 
sequential decay pattern.

\subsubsection{Two-proton emission from $^{17}${\rm Ne}}

\begin{figure}[hht]
\begin{center}
\includegraphics[scale=0.5,angle=-90]{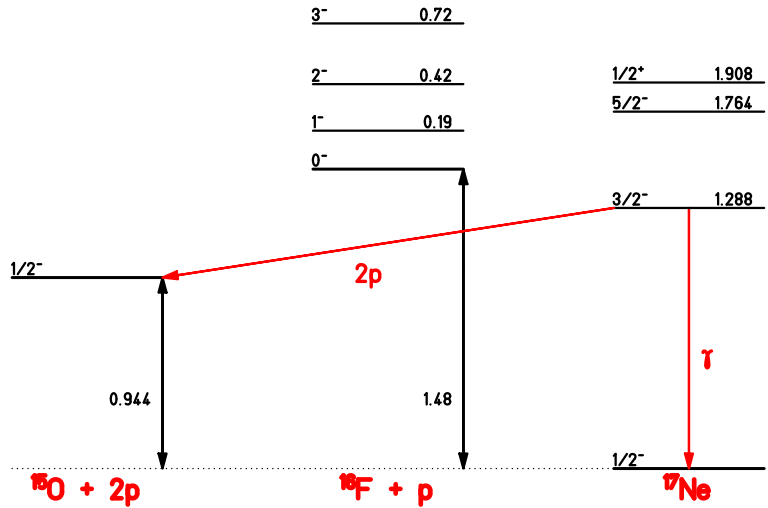} 
\end{center}
\caption{Decay scheme of the excited states in $^{17}$Ne. The first excited state 
         in $^{17}$Ne is bound with respect to one-proton emission but unbound with 
         respect to two-proton emission to the ground state of $^{15}$O. Hence, direct 
         2p decay is expected. For further details see the discussion in the text.}
\label{fig:ne17_scheme}
\end{figure}

The decay of excited states of $^{17}$Ne by 2p emission was studied several 
times in recent years~\cite{chromik97,chromik02,zerguerras04}. The first
excited state of $^{17}$Ne is 1p bound but unbound with respect to 2p
decay by 344~keV and may therefore decay by direct 2p emission to the
ground state of $^{15}$O (see Fig.~\ref{fig:ne17_scheme}). Chromik et 
al.~\cite{chromik97,chromik02} populated the first excited states of $^{17}$Ne by 
Coulomb excitation and observed their decay by $\gamma$ de-excitation~\cite{chromik97} or
in a complete-kinematics experiment~\cite{chromik02}.

In the first experiment, the missing strength of the decay of the first excited
state by $\gamma$ emission, as compared to theoretical calculations, was interpreted 
as a possible unobserved 2p branch. However, it was found that such 
an interpretation was in contradiction to barrier-penetration calculations. 
In fact, a 2p partial half-life shorter by a factor 1700 than predicted 
was needed to account for the missing decay strength.
The second complete-kinematics experiment allowed determining all decay channels 
and a partial 2p half-life for the first excited state larger than 26~ps was deduced.
No 2p emission from the first excited state was observed. The decay of the 
second excited state, also populated by Coulomb excitation, was observed to decay by 
sequential 2p emission to the ground state of $^{15}$O yielding an isotropic
emission pattern for the two protons.

This experimental result was confirmed by Zerguerras et al.~\cite{zerguerras04}.
In their experiment, excited states of $^{17}$Ne were populated by one-neutron
stripping reactions from a $^{18}$Ne beam at 36~MeV/nucleon. States up to 10~MeV excitation 
energy were observed to decay by 2p emission. The decay products were
detected in the MUST detector array~\cite{blumenfeld99} and in the SPEG 
spectrometer~\cite{speg}. These measurements allowed a reconstruction of
the complete decay kinematics and therefore to determine the excitation energy 
of $^{17}$Ne and the relative proton-proton angle in the center-of-mass frame.

\begin{figure}[hht]
\begin{center}
\includegraphics[scale=0.5,angle=-0]{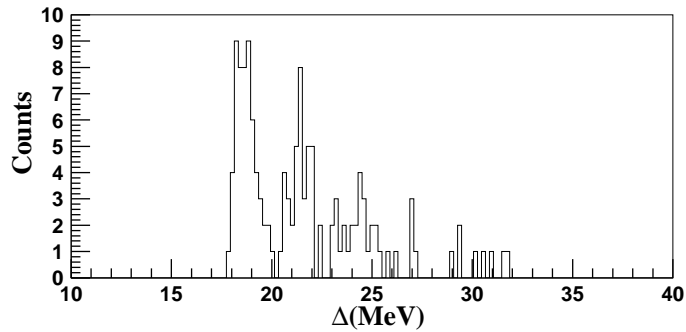} 
\end{center}
\caption{Invariant mass spectrum for $^{15}$O + 2p events~\cite{zerguerras04}. The peak at 18.5~MeV
         is interpreted as the decay of the second and third excited states 
         of $^{17}$Ne by two-proton emission, whereas the activity above 20~MeV 
         arises from the decay of higher-lying states.}
\label{fig:zerg1}
\end{figure}

\begin{figure}[hht]
\begin{center}
\includegraphics[scale=0.5,angle=-90]{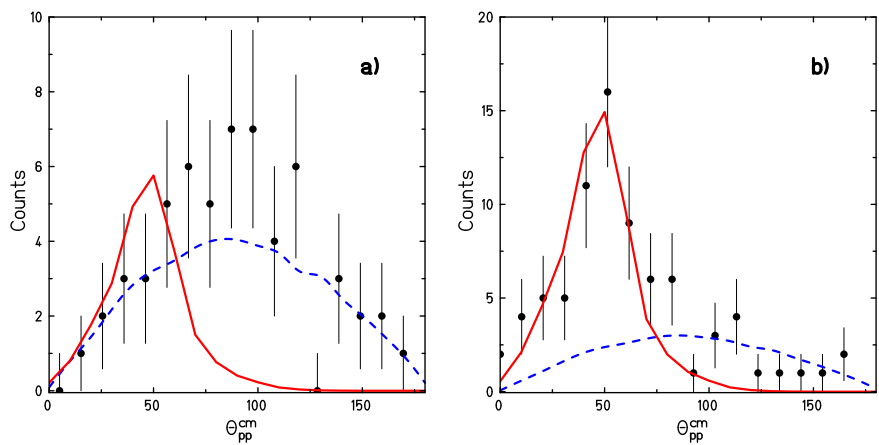} 
\end{center}
\caption{Proton-proton angular distribution for 2p events with a $^{15}$O recoil 
         yielding a mass excess of less than 20.2~MeV (a) and above 20.2~MeV (b) 
         (see Fig.~\ref{fig:zerg1}). The solid and the dashed curves
         are simulations of a correlated two-proton emission
         via a $^2$He resonance and a sequential emission pattern, respectively~\cite{zerguerras04}.}
\label{fig:zerg2}
\end{figure}

Fig.~\ref{fig:zerg1} shows the invariant mass spectrum for $^{15}$O+2p events.
The angular distribution of the two protons was measured to be isotropic for the 
second and third excited states in $^{17}$Ne (Fig.~\ref{fig:zerg2}a). However, for higher
lying states, a distinct angular correlation of the two protons could be observed for 
the first time (Fig.~\ref{fig:zerg2}b). This result is rather surprising, as many
intermediate states in the 1p daughter nucleus $^{16}$F should be accessible and therefore
a sequential decay mechanism via these states should dominate.

The interpretation of this decay pattern is still controversial. Kanungo et al.~\cite{kanungo03} and Grigorenko et 
al.~\cite{grigorenko05} suggested that some of the excited states of $^{17}$Ne may have a pronounced 2p halo structure and
a large overlap with the $^{15}$O ground state leading to much larger spectroscopic factors for a direct 2p decay than 
for a sequential decay.  Obviously, the validity of this assertion
depends on whether the configuration mixing in $^{17}$Ne leaves the $^{15}$O core intact. 
On the other hand, as shown in SMEC studies~\cite{rotureau05} 
the correlated 2p emission is {\em always} hindered if 1p decay channels are open. Moreover, the measured energies 
of the two emitted protons from $^{17}$Ne show large differences which cannot be reconciled with a diproton scenario 
where roughly equal energies are expected for the two protons. 

Another explanation of the 2p emission pattern in $^{17}$Ne could be the large deformation in higher lying states. 
Strong anisotropy of the Coulomb barrier could be a source of dynamical correlations between emitted protons 
even if this 2p decay is a sequence of two successive 1p emissions. Effects of this kind have been discussed 
in the de-excitation of heavy-ion residues (see e.g. Ref.~\cite{lefevre04}). A similar explanation has been put forward 
also in connection with the anomalous decay properties of  $^{94}$Ag$^m$~\cite{mukha06}. Obviously, for a consistent 
interpretation of the  $^{15}$O + 2p experiment, higher statistics data is mandatory.

\subsubsection{Two-proton emission from $^{18}${\rm Ne}}

In an experiment performed at ORNL Oak Ridge, Gomez del Campo et 
al.~\cite{gomez01} used a radioactive $^{17}$F beam on a (CH$_2$)$_n$ target to
produce excited states in $^{18}$Ne. The experiment used a thick-target approach, i.e. only 
the protons could be detected behind the target, whereas the beam-like particles (e.g.
the heavy $^{16}$O recoils from the decay of excited states in $^{18}$Ne) could not 
penetrate the target. Therefore, only protons with a relatively large energy loss depending
on their emission point in the target could be observed. Using two different incident energies of the 
$^{17}$F beam, these authors observed at higher incident energy the emission of two 
protons which they interpreted as a decay of the 6.15~MeV resonance state in $^{18}$Ne.
For the lower energy, where this state could not be populated, no coincident protons were
observed. Both the relative proton-proton angle and the energy difference between the
two protons was in agreement with simulations for a three-body decay of this state as well
as for a diproton-type decay. This lack of sensitivity is to a large extent due to
the limited statistics of the experiment, but also due to the limited angular
coverage of the experimental setup.

\subsubsection{Two-proton emission from $^{94}${\rm Ag}}

The decay of $^{94}$Ag$^m$ is unusual in several respects. It possesses two $\beta$-decaying 
isomers and has the highest spin state ever observed which decays by $\beta$ decay. 
In particular, the second isomeric level seems 
to decay by several different decay branches: $\beta$-delayed $\gamma$ decay~\cite{lacommara02,plettner04}, 
$\beta$-delayed proton emission~\cite{mukha04}, direct proton 
decay~\cite{mukha05}, and direct 2p emission~\cite{mukha06}. All these experiments were performed at the 
on-line separator of GSI~\cite{roeckl06}.

The two protons were observed by means of silicon strip detectors which surrounded the implantation point 
of the on-line separator. This setup allowed for the measurement of the individual proton energies and the 
relative proton-proton angle. Although the statistics of the 2p experiment is rather low~\cite{mukha06}, 
the authors identified proton-proton correlations. They interpreted these data as due to simultaneous 2p 
emission from a strongly deformed ellipsoidal nucleus. In this case, the asymmetric Coulomb barrier favours 
the emission of two protons within narrow cones around the poles of the ellipsoid, either from the same pole 
or on opposite sides. 

One should mention, that  the observed proton-proton correlations  in such a scenario are largely unrelated 
to the proton-proton correlations inside a nucleus. The microscopic theoretical study of the $^{94}$Ag$^m$ decay 
is also impossible using the present-day computers, though only such studies could disentangle an internal 
structure of decaying states from dynamical effects of the anisotropic Coulomb barrier, providing more details 
about the 2p decay pattern in the nucleus. 
Unfortunately, further experimental studies of the $^{94}$Ag$^m$ decay cannot be made without  new technical 
developments to produce $^{94}$Ag$^m$, as the on-line separator of GSI has been dismantled a few years ago.

\section{Two-proton radioactivity}

Goldanskii~\cite{goldanskii60,goldanskii62} called it a pure case of 2p radioactivity, if 
1p emission is not possible because the 2p emitter level is lower in energy than 
the level in the 1p daughter nucleus and, moreover, the corresponding levels are sufficiently 
narrow and thus do not overlap. The limitations of this picture will be discussed in chapter 8. 
Goldanskii also gave an arbitrary half-life limit of 10$^{-12}$~s~\cite{goldanskii88} to distinguish between 
two-proton radioactivity and the decay of particle-unstable nuclei. 
None of the examples discussed above satisfy Goldanski's radioactivity criterion; 
they all have rather short half-lives (of the order of 10$^{-21}$~s) and
for the light ground-state emitters, the states which emit the first and the second proton 
overlap significantly. In this sense, 2p radioactivity was observed for the first time 
in  $^{45}$Fe experiments~\cite{giovinazzo02,pfuetzner02}. Later $^{54}$Zn~\cite{blank05zn54} and 
possibly $^{48}$Ni~\cite{dossat05} were shown to be other ground-state 2p emitters. 

Recent theoretical predictions~\cite{brown91,ormand96,cole96} agree that $^{39}$Ti, 
$^{42}$Cr, $^{45}$Fe, and $^{49,48}$Ni are candidates to search for 2p radioactivity.
Firstly, their predicted decay energy is close to or in the range of energies (1.0-1.5~MeV) where
2p radioactivity is most likely expected to occur, and secondly these isotopes are already or
could soon be produced in projectile-fragmentation type experiments. This production technique allows
to unambiguously identify the isotopes before their decay and thus measure the decay half-life and the
decay energy, the necessary ingredients to identify this new radioactivity. The most promising 
candidates were $^{45}$Fe and $^{48}$Ni with decay energies of 1.1-1.4~MeV.

\subsection{Identification of the 2p candidates and first negative results}

The first of these isotopes, $^{39}$Ti, was observed at the 
Grand Acc\'el\'erateur National Ions Lourds (GANIL),
Caen~\cite{detraz90} in 1990. This 
experiment was also the first attempt to observe 2p radioactivity of these medium-mass
nuclei after projectile fragmentation. However, no indication for 2p radioactivity could be
found. Instead, a $\beta$-decay half-life of 28$^{+8}_{-7}$ms was measured and $\beta$-delayed 
protons were observed. Another attempt to search for 2p radioactivity from $^{39}$Ti was made 
at GANIL  in 1991 with an improved setup~\cite{faux94,blank93mu}. However, again no 2p events could be observed. 

The nuclei $^{42}$Cr, $^{45}$Fe, and $^{49}$Ni were identified for the first time a few years later at 
GSI~\cite{blank96fe45}. In this experiment, no spectroscopic studies could be performed 
and the observation of these isotopes was the only outcome. Nonetheless, this experiment provided 
the first observation of one of the most promising 2p radioactivity candidates and opened the door for the
search of $^{48}$Ni, another candidate. This doubly-magic nucleus was then discovered in 1999~\cite{blank00ni48} in a 
GANIL experiment using the SISSI/LISE3 device~\cite{lise}. In the same experiment, the 
observation of $^{45}$Fe was confirmed with about a factor of 10 higher statistics. 

The GANIL experiment was also set up to perform spectroscopic studies of the very exotic 
nuclei produced~\cite{giovinazzo01}. The decay trigger in this experiment allowed mainly 
$\beta$-delayed events
to be registered. $\beta$-delayed decay events from e.g. $^{45}$Fe decay had a detection efficiency of
about 30\%, as they were triggered by two detectors adjacent to the implantation detector.
2p events without $\beta$ particle had a very low trigger rate (a few percent) due to a
heavy-ion trigger for the implantation device. This situation allowed for the observation of
the decay of $^{45}$Fe, however, basically of only its $\beta$-delayed branch (see Sect. 6.2). 
In any case, no claim about the observation of 2p radioactivity was made.

\begin{figure}[hht]
\begin{minipage}{7.5cm}
\includegraphics[scale=0.65,angle=0]{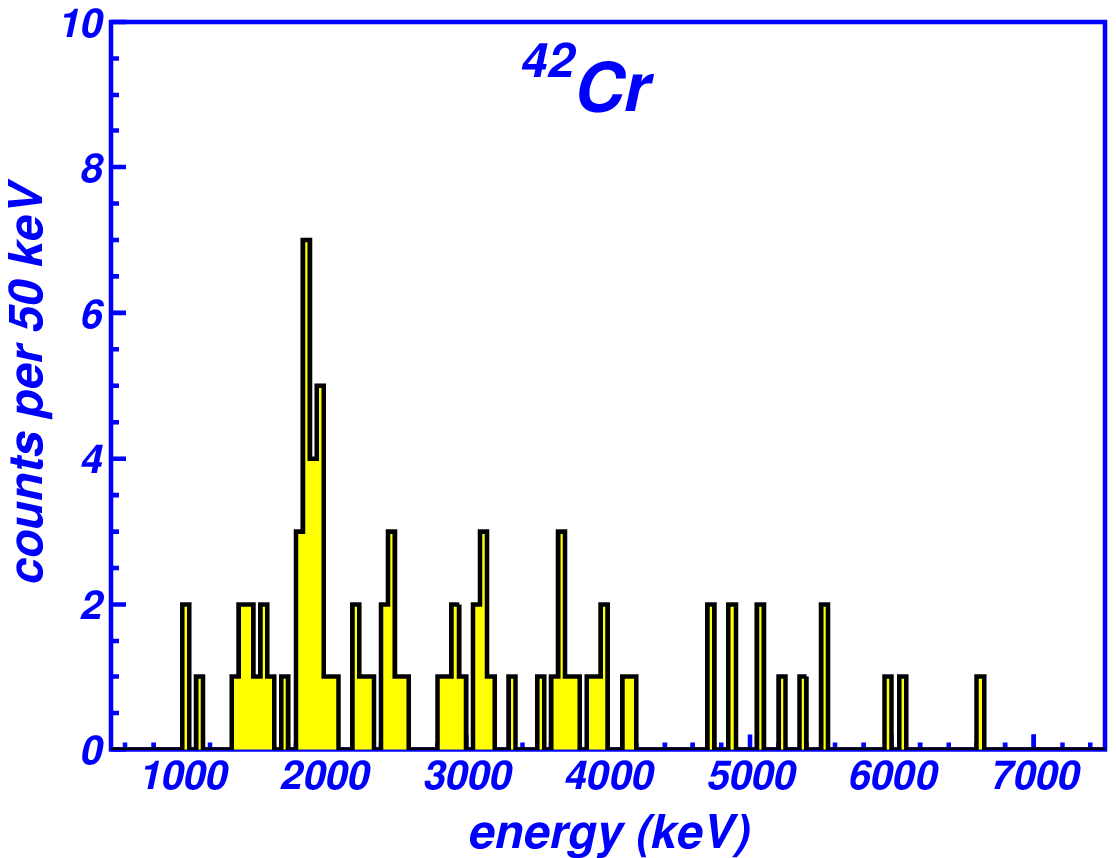}
\end{minipage} 
\ \hfill \
\begin{minipage}{7.5cm}
\includegraphics[scale=0.4,angle=-90]{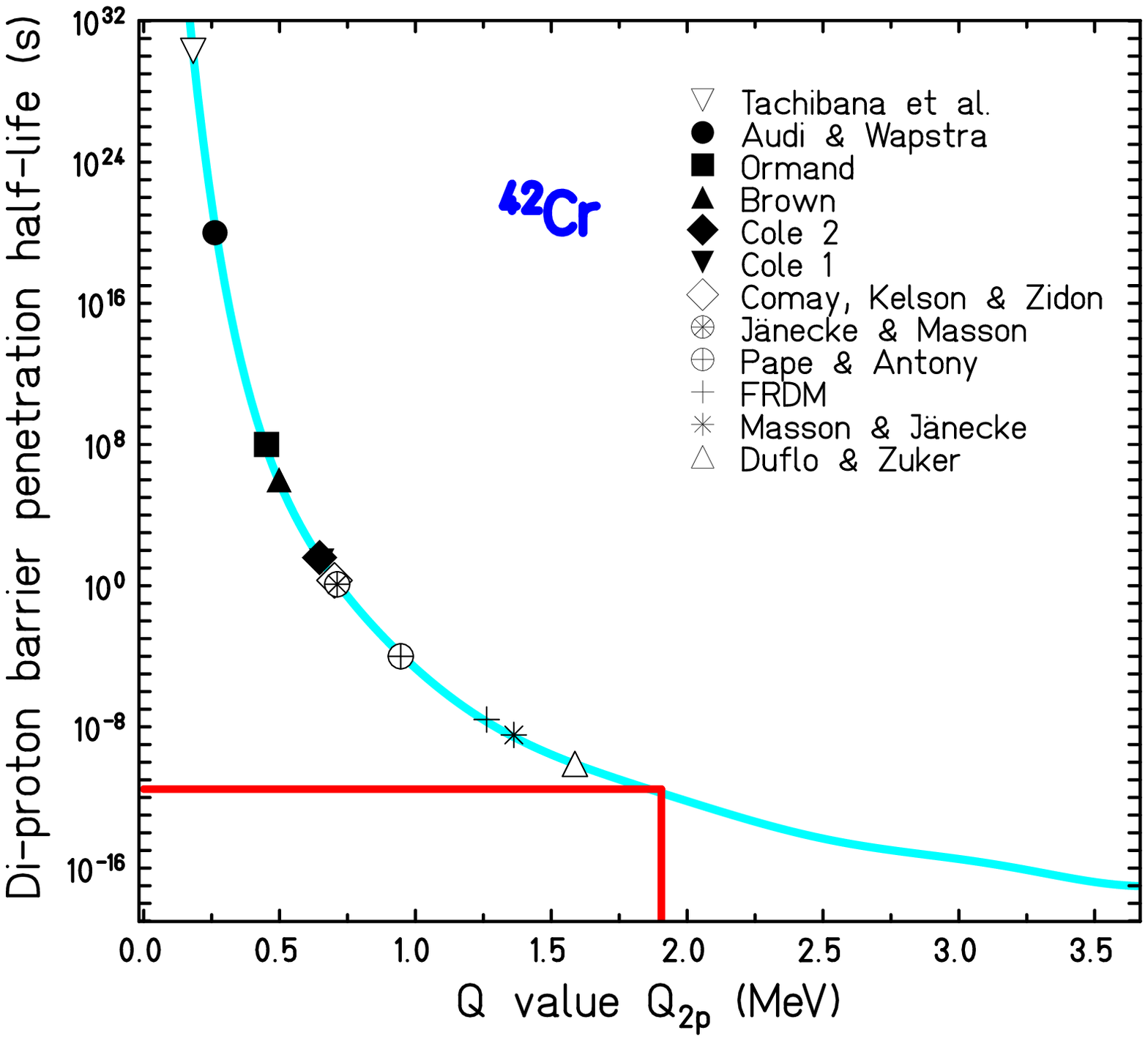} 
\end{minipage} 
\caption{Left-hand side: Charged-particle decay-energy spectrum obtained after $^{42}$Cr implantation.
         The 1.9~MeV peak is interpreted as a $\beta$-delayed decay~\cite{giovinazzo01}. Right-hand side: Barrier penetration
         half-life for a di-proton in $^{42}$Cr as a function of the available decay 
         energy~\cite{giovinazzo01}. The red line represents the 1.9~MeV peak from the decay-energy spectrum
         yielding a barrier penetration half-life of 10$^{-12}$s if it were of 2p origin.}
\label{fig:cr42}
\end{figure}

Nevertheless, the experiment allowed one to test two other possible, although less promising, 2p candidates:  
$^{42}$Cr and $^{49}$Ni. As mentioned above, these two nuclei were predicted to be possible 2p emitters, 
however, with decay energies most likely too small to exhibit this decay. The decay energy spectrum obtained
during this experiment for $^{42}$Cr is shown in Fig.~\ref{fig:cr42}. The prominent peak at 1.9~MeV
was excluded to be of 2p nature, as such a high energy would lead to an extremely short barrier penetration
time for the two protons (see Fig.~\ref{fig:cr42}, right-hand side), in contradiction with the $^{42}$Cr 
half-life of 13.4$^{+3.6}_{-2.4}$ms measured in  the same experiment.

In a similar way, $^{49}$Ni could be excluded from having a significant direct 2p branch. The last
2p emitter candidate tested in this experiment was $^{39}$Ti. In agreement with previous experiments, 
this nucleus turned out to decay predominantly by $\beta$ decay. Several $\beta$-delayed proton groups
could be identified and the 2p emission Q value (Q$_{2p}$ = 670$\pm$100 keV) could be determined, 
indicating a rather high barrier-penetration half-life and therefore a rather low probability for 
2p radioactivity.

\subsection{The discovery of two-proton radioactivity}

\begin{figure}[hht]
\begin{minipage}{7.5cm}
\includegraphics[scale=0.8,angle=0]{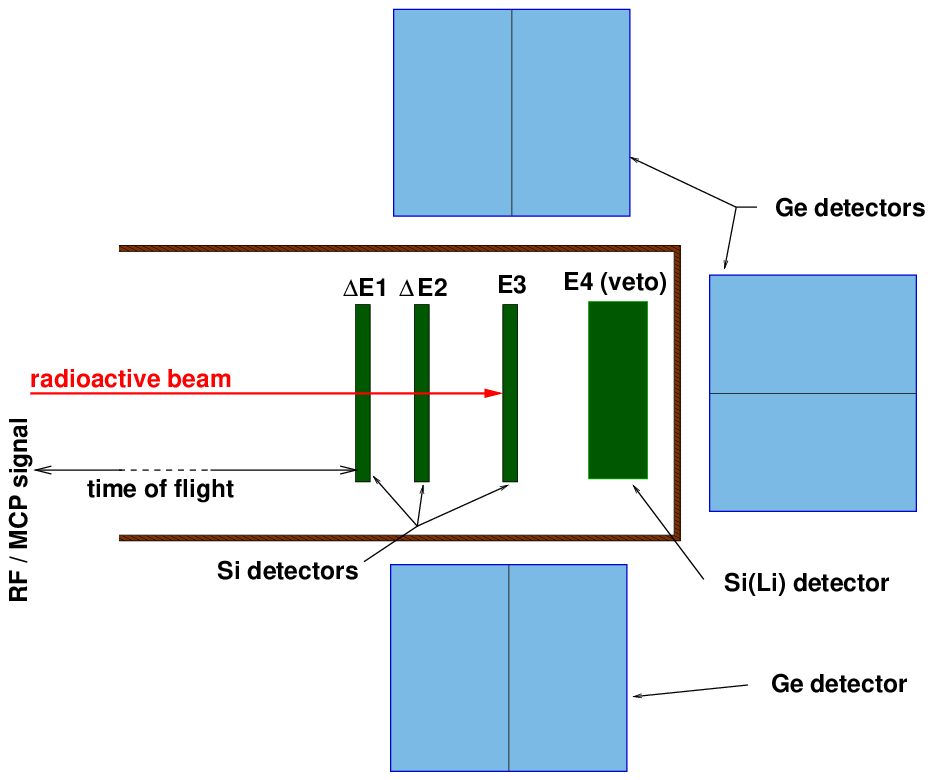}
\end{minipage} 
\ \hfill \
\begin{minipage}{7.5cm}
\includegraphics[scale=0.41,angle=-0]{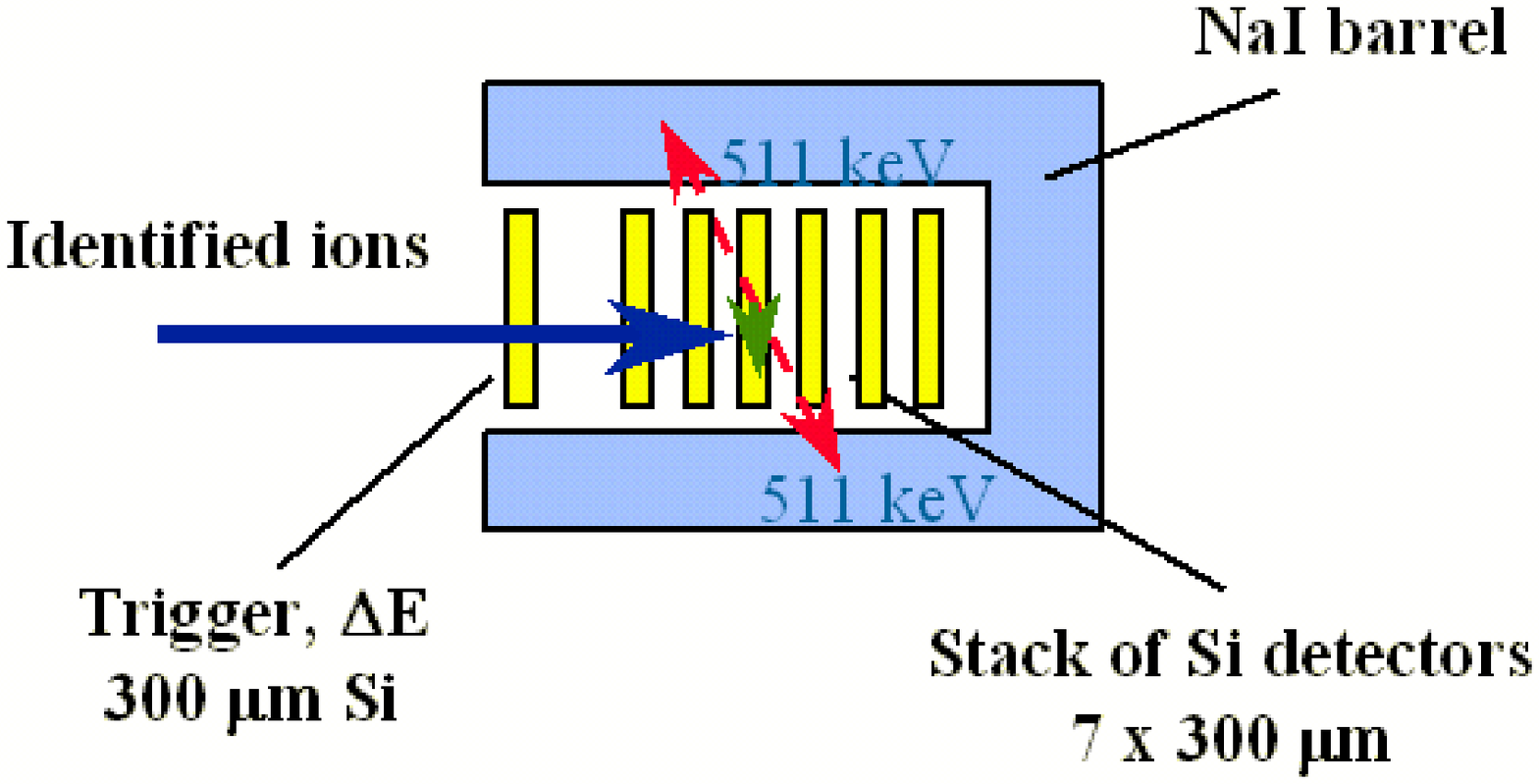} 
\end{minipage} 
\caption{Left-hand side: Detection setup used in the GANIL experiment to identify two-proton  radioactivity. 
         The setup consisted of a silicon telescope, where the implantation device was a double-sided
         silicon strip detector (DSSSD). Several time-of-flight, energy-loss and residual energy measurements
         allowed an unambiguous identification of the projectile fragments. The radioactive decay of 
         the isotopes of interest was detected in the DSSSD. The last detector, a Si(Li) detector, was used
         to detect $\beta$ particles~\cite{giovinazzo02}. Right-hand side: Setup used in the GSI experiment.
         A silicon telescope of seven elements was surrounded by a high-efficiency NaI array.
         Implantation and decay events were correlated in different detectors depending of the
         implantation depth of the $^{45}$Fe ions. The NaI device served to search for $\gamma$ radiation
         from positron annihilation~\cite{pfuetzner02}.}
\label{fig:fe45_setup}
\end{figure}

The discovery of ground-state 2p radioactivity had to await the year 2002, when two experiments, 
one performed in 2000 at the SISSI/LISE3 facility of GANIL~\cite{giovinazzo02} and one performed at the FRS
of GSI in 2001~\cite{pfuetzner02}, were published. Both experiments used the fragmentation of a
$^{58}$Ni primary beam. The fragments of interest were implanted in a silicon detector telescope 
(see Fig.~\ref{fig:fe45_setup}), where correlations in time and space between implantation events 
and decay events could be performed. This technique allowed to unambiguously correlate $^{45}$Fe 
implantation events with their decays. 

\begin{figure}[hht]
\begin{minipage}{7.5cm}
\includegraphics[scale=0.90,angle=0]{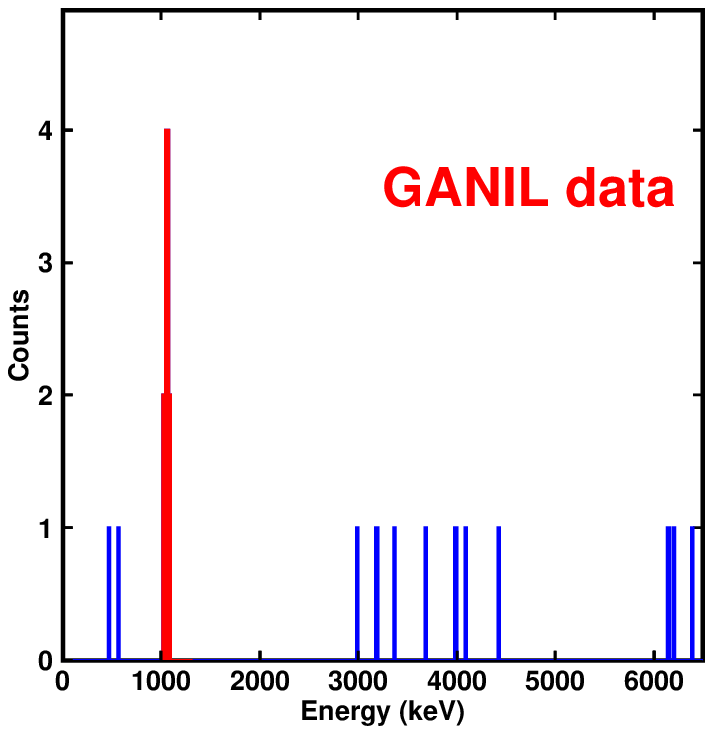}
\end{minipage} 
\ \hfill \
\begin{minipage}{7.5cm}
\includegraphics[scale=0.90,angle=-0]{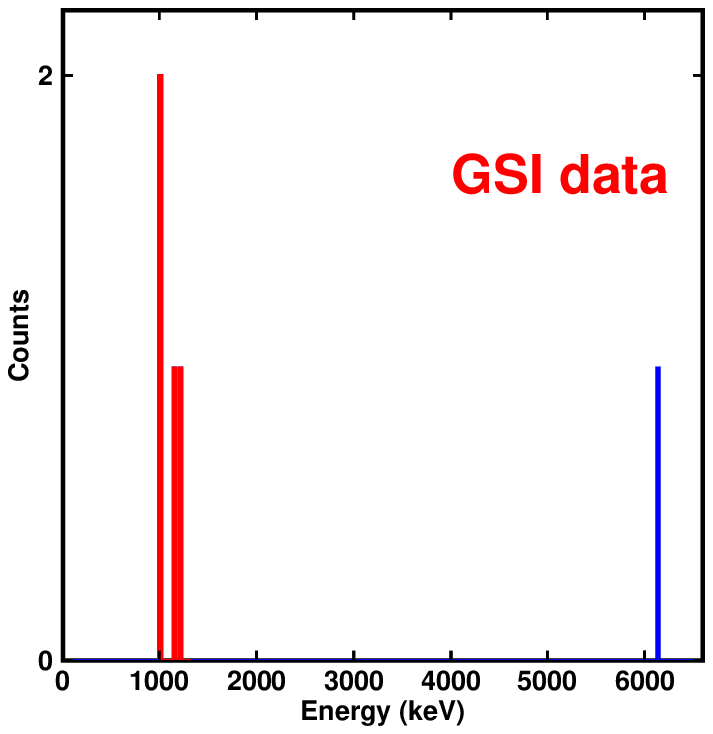} 
\end{minipage} 
\caption{Left-hand side: $^{45}$Fe decay-energy spectrum from the GANIL experiment exhibiting a peak at
         (1.14$\pm$0.04)~MeV~\cite{giovinazzo02}. Right-hand side: $^{45}$Fe decay-energy spectrum from 
         the GSI experiment showing four events at 1.1(1)~MeV~\cite{pfuetzner02}. Both spectra 
         also show events at higher energies.}
\label{fig:fe45_e}
\end{figure}

Fig.~\ref{fig:fe45_e} shows the decay energy spectra from the two experiments. Both find a decay 
energy of about 1.1~MeV with a decay half-life in the 3-5~ms range. An additional feature of both experiments
was the capability to detect additional radiation like $\beta$ particle in the GANIL experiment, or
511~keV annihilation radiation and other $\gamma$ rays in the GSI experiment (see Fig.~\ref{fig:fe45_setup}).
This capability allowed for a rejection with a high probability of a possible $\beta$-delayed origin 
of the decay events with an energy release of about 1.1~MeV. Other pieces of evidence from the GANIL experiment
were the smaller width of the 1.14~MeV peak as compared to $\beta$-delayed proton emission peaks in
neighbouring nuclei and the observation of daughter decays consistent with the decay of $^{43}$Cr, 
the 2p daughter of $^{45}$Fe. In addition, the experimentally observed decay energy is in nice agreement 
with theoretical predictions~\cite{brown91,ormand96,cole96}. These experimental results led the authors 
to conclude the discovery of ground-state 2p radioactivity in the decay of $^{45}$Fe.

Of particular interest was also the observation of decay events from $^{45}$Fe with an energy 
release different from the 1.1~MeV of the main peak. Some of these additional events could be identified
as being in coincidence with $\beta$ particles and were therefore attributed to $\beta$-delayed
decay branches. From these first results, a 2p radioactivity branching ratio of 70-80\% was determined
for $^{45}$Fe. The partial half-lives of $^{45}$Fe for 2p emission and for $\beta$ decay are therefore
of the same order and allowed the two decay modes to compete.

These results on $^{45}$Fe have recently been confirmed in a new GANIL experiment~\cite{dossat05}.
Several experimental improvements allowed a better energy calibration and a smaller data acquisition
dead time. The decay energy was determined to be 1.154(16)~MeV, the half-life to be 1.6$^{+0.5}_{-0.3}$ms, 
and the 2p branching ratio to be 57(10)\%. Together with the results from the two earlier 
experiments~\cite{giovinazzo02,pfuetzner02}, this experiment allowed to determine average values
for the decay energy (Q$_{2p}$= 1.151(15)~MeV), the half-life (T$_{1/2}$= 1.75$^{+0.49}_{-0.28}$ms), 
the 2p branching ratio (BR~= 59(7)\%), and the partial 2p half-life (T$_{1/2}^{2p}$= 3.0$^{+0.9}_{-0.6}$ms). 

In 2007, a new experiment performed at the A1900 separator of Michigan State University allowed the observation of
2p events in a TPC (see below). The newly determined half-life and the 2p branching ratio are included in 
Table~\ref{tab:2p-emitters} to determine average values.
These average values will be used in the comparison with theoretical model of 2p radioactivity.

\begin{figure}[hht]
\begin{center}
\includegraphics[scale=0.5,angle=-90]{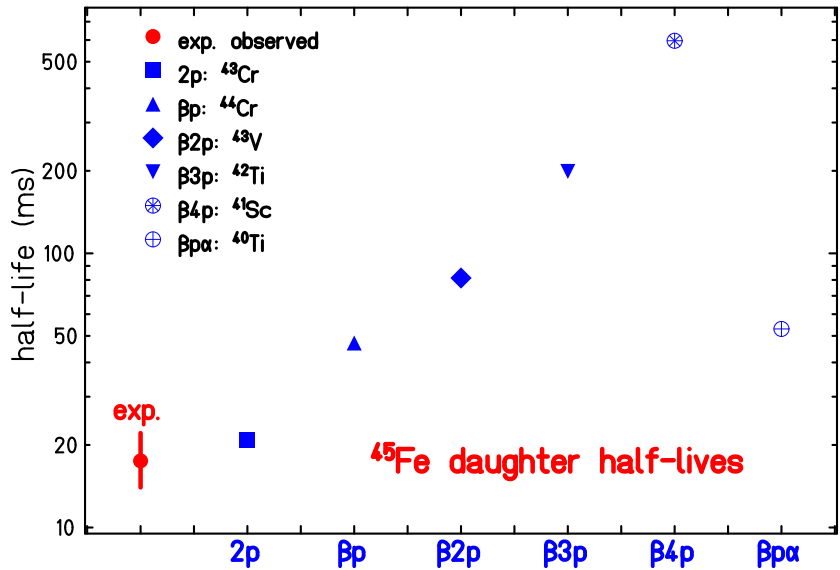}
\caption{The experimentally observed half-life for the decay of the daughter of $^{45}$Fe is compared to the 
         half-lives of all possible daughters~\cite{dossat07}. Only the half-life of $^{43}$Cr is
         in agreement with the experimental value.}
\label{fig:daughter_half-life}
\end{center}
\end{figure}

Another outcome from the GANIL experiments~\cite{giovinazzo02,dossat05} on $^{45}$Fe was the determination 
of the half-life of the 2p daughter nucleus. From the observation of the second decay after $^{45}$Fe implantation, the 
half-life of the daughter decay, in coincidence with the 1.151~MeV peak, could be determined. 
In Fig.~\ref{fig:daughter_half-life}, this half-life is compared to the half-lives 
of all possible $^{45}$Fe daughter nuclei. Only the half-life of $^{43}$Cr is
in agreement with the experimentally observed daughter half-life, thus giving an independent proof for the 
observation of 2p radioactivity.

\subsection{Two-proton radioactivity of $^{54}${\rm Zn}}

The nucleus $^{54}$Zn was early identified as a possible 2p emitter~\cite{goldanskii61,goldanskii62}.
More recent $Q_{2p}$ estimates yielded decay energies of 1.79(12)~MeV~\cite{cole96}, 
1.87(24)~MeV~\cite{ormand97}, and 1.33(14)~MeV~\cite{brown02a}. Due to rather large
theoretical error bars and the uncertainty concerning the decay mechanism, it was 
rather unclear whether $^{54}$Zn would exist at all, i.e. whether its half-life would 
be long enough to be observed in projectile-fragmentation type experiments. For this 
purpose, a life time of the order of a few hundred nano-seconds was mandatory.

The search for $^{54}$Zn began with the observation of two other proton-rich zinc isotopes, 
$^{56}$Zn and $^{55}$Zn~\cite{giovinazzo01a}. They were observed for the first time in experiments
at the LISE3 separator~\cite{lise} via 2p pick-up reactions with a $^{58}$Ni primary beam. The identification
of these two isotopes allowed also for an extrapolation of the production cross section of $^{54}$Zn.

\begin{figure}[hht]
\begin{center}
\includegraphics[scale=0.4,angle=-0]{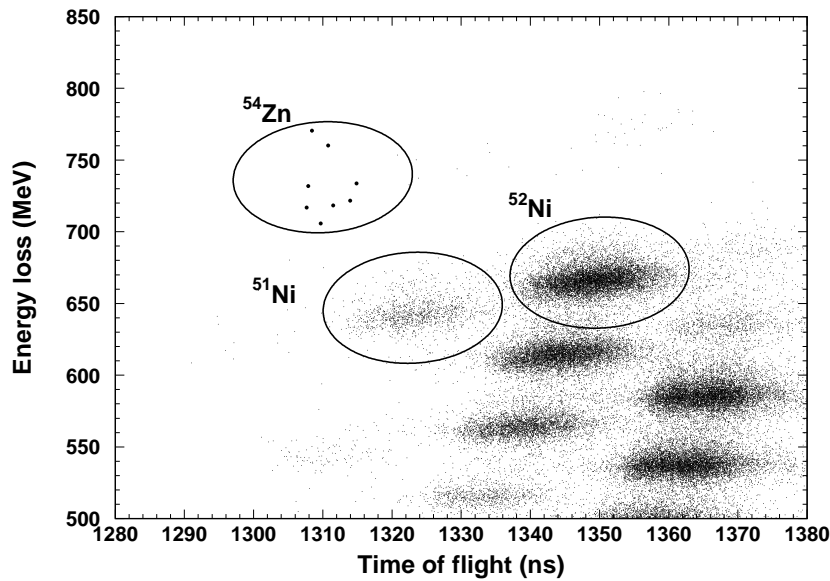} 
\end{center}
\caption{Two-dimensional identification plot for $^{54}$Zn and neighbouring nuclei. The energy loss
         in the first silicon detector of the detection setup is plotted as a function of the time-of-flight
         through the separator. Additional parameters are used to purify the spectrum~\cite{blank05zn54}.}
\label{fig:zn54_id}
\end{figure}

The nucleus $^{54}$Zn was synthesized and observed for the first time in 2004 in an experiment at the
SISSI/LISE3 facility of GANIL Caen~\cite{blank05zn54}. Similarly to the experiment with 
$^{45}$Fe, $^{54}$Zn
was produced with a primary $^{58}$Ni beam at 75~MeV/nucleon impinging on a natural nickel target.
The fragments of interest were selected and separated with the LISE3 separator and implanted 
in a silicon telescope with a double-sided silicon strip detector  being the central device. The $^{54}$Zn nuclei were identified
by means of time-of-flight, energy-loss and residual-energy measurements. Fig.~\ref{fig:zn54_id}
shows the two-dimensional identification plot for the $^{54}$Zn experiment. All in all, eight $^{54}$Zn
nuclei could be unambiguously identified.

\begin{figure}[hht]
\begin{center}
\includegraphics[scale=0.5,angle=-0]{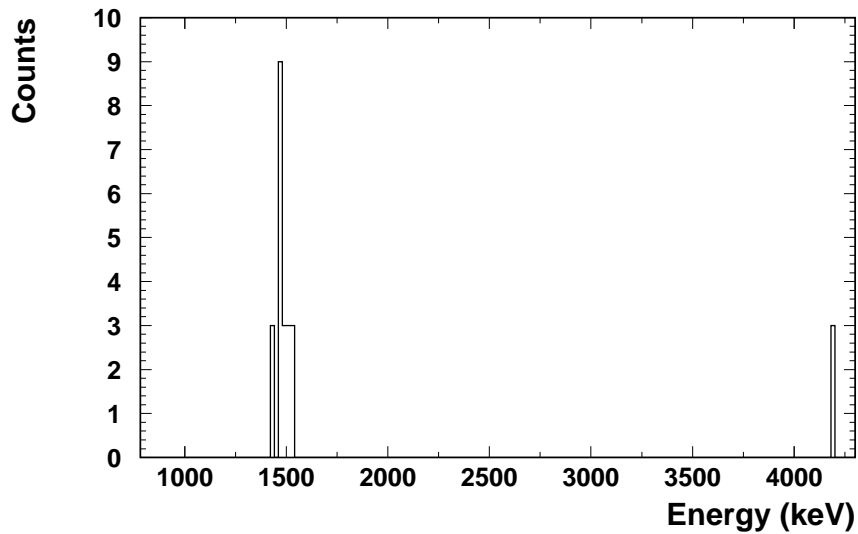} 
\end{center}
\caption{Charged-particle spectrum from the decay of $^{54}$Zn. The peak at 1.48(2)~MeV is due
          to two-proton emission from the ground state of $^{54}$Zn. The event at 4.19~MeV is in
          coincidence with a $\beta$ particle in the adjacent detector~\cite{blank05zn54}.}
\label{fig:zn54_e}
\end{figure}

The setup allowed correlating in time and space these implantation events with subsequent decays.
Seven of the eight implantation events are followed by a decay with an energy release of 1.48(2)~MeV
(see Fig.~\ref{fig:zn54_e}).
None of these decay events is in coincidence with a $\beta$ particle. As for $^{45}$Fe, the 1.48~MeV peak
is much narrower than $\beta$-delayed proton lines in neighbouring nuclei. Although less significant
as in the case of $^{45}$Fe, daughter decays in agreement with the decay of $^{52}$Ni, the 2p daughter 
of $^{54}$Zn, could be observed. Finally, the experimental half-life (T$_{1/2}$= 3.2$^{+1.8}_{-0.8}$ms)
is in agreement with what is expected for a 2p decay with an energy release of 1.48~MeV. From these
experimental observations, the authors concluded the observation of ground-state 2p
radioactivity from $^{54}$Zn. The 2p branching ratio was determined to be 87$^{+10}_{-17}$\%,
which yielded a partial half-life for 2p decay of 3.7$^{+2.2}_{-1.0}$~ms. These results will be compared
to theoretical predictions in Sect. 9.3.

\subsection{The decay of $^{48}${\rm Ni}}

The discovery of $^{48}$Ni and the observation of its decay was considered important for several reasons. 
Firstly, it is a doubly-magic nucleus which was predicted to be particle stable or quasi-stable
only due to strong shell effects. Secondly,  $^{48}$Ni  is one of the most exotic  of
all doubly-magic nuclei within experimental reach. Its observation and the measurement of its properties, 
like e.g. the excitation energy of the first 2$^+$ state, were expected to teach us a lot about the 
persistence of shell structure far from the valley of stability. 

The nucleus $^{48}$Ni was rather early identified as a possible 2p emitter~\cite{goldanskii60}. This prediction
was confirmed and refined by new mass and $Q_{2p}$ calculations, which yielded 1.36(13)~MeV~\cite{brown91}, 
1.14(21)~MeV~\cite{ormand96}, 1.35(6)~MeV~\cite{cole96}, and 
1.29(33)~MeV~\cite{ormand97}. When compared to $^{45}$Fe and $^{54}$Zn, it is evident that this
$Q_{2p}$ value is again in the right range to permit $^{48}$Ni to have a significant 2p radioactivity decay branch.

\begin{figure}[hht]
\begin{center}
\includegraphics[scale=0.4,angle=-0]{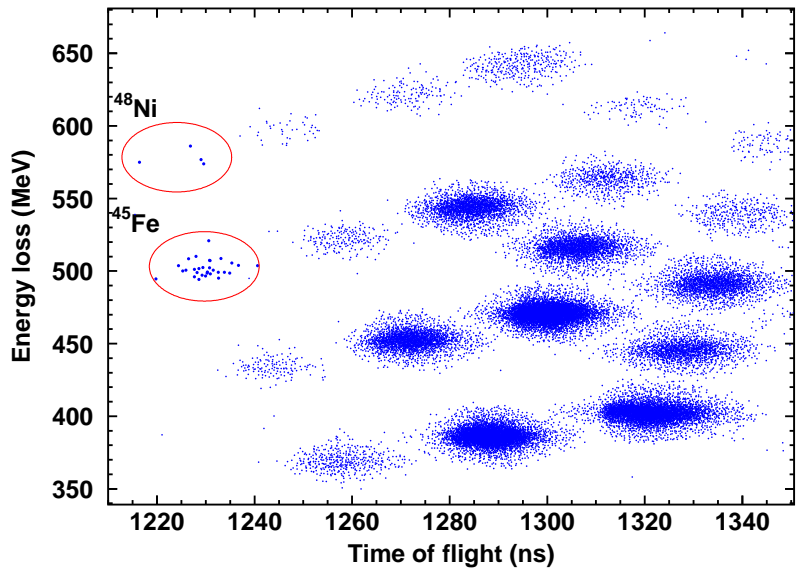} 
\end{center}
\caption{Two-dimensional identification plot for $^{48}$Ni and neighbouring nuclei. The energy loss
         in the first silicon detector is plotted as a function of the flight time
         through the separator. Additional parameters are used to purify the spectrum~\cite{dossat05}.}
\label{fig:ni48_id}
\end{figure}

The isotope $^{48}$Ni was first observed in 1999 at GANIL Caen~\cite{blank00ni48}, where four events could be 
unambiguously identified. This result was confirmed in 2004, when again four events of $^{48}$Ni could be registered
in a new SISSI/LISE3 experiment at GANIL~\cite{dossat05}. These data were obtained in the same run
as the second $^{45}$Fe data set from GANIL,  because the acceptance of the SISSI/LISE3 device allows the measurement 
of different neighbouring nuclei at the same time. Fig.~\ref{fig:ni48_id} shows the fragment identification plot.

\begin{figure}[hht]
\begin{center}
\includegraphics[scale=0.7,angle=-0]{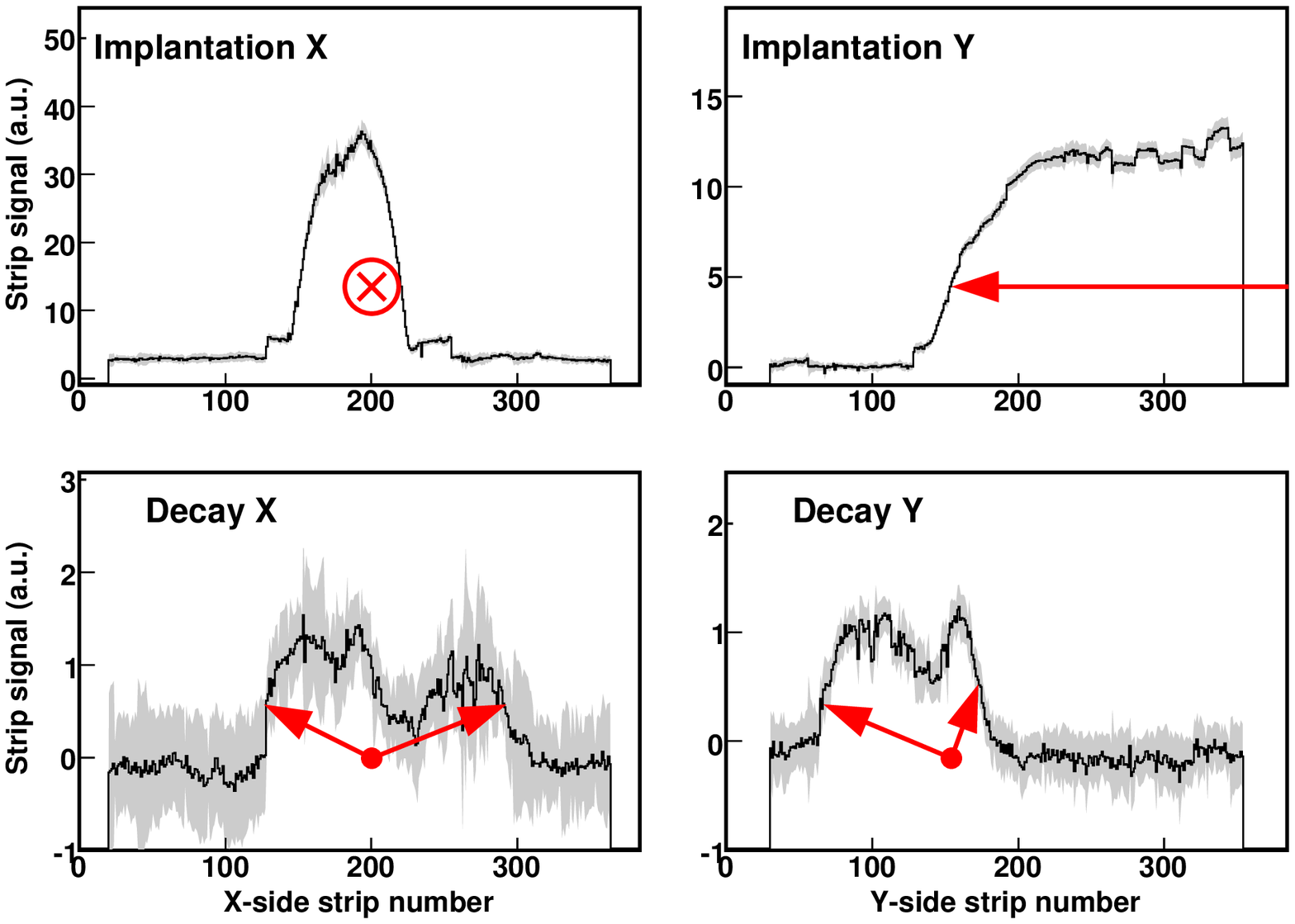} 
\end{center}
\caption{A single event of two-proton decay of $^{45}$Fe from a recent experiment performed with the Bordeaux TPC 
         at the LISE3 facility of GANIL~\cite{giovinazzo07}. The top row shows the $^{45}$Fe implantation event, where the ion enters 
         the chamber parallel to the X direction and stops in the center of the chamber. The decay (bottom) takes 
         place at the point where the ion stopped.}
\label{fig:2pevent}
\end{figure}

The four $^{48}$Ni implantation events could be correlated in time and space with subsequent 
decay events. From the four decay events, only one event has all characteristics of a 2p emission: 
a decay energy of 1.35(2)~MeV in the expected energy range and no coincident $\beta$ particle. The decay takes place after 
1.66~ms and the daughter decay is also fast as expected for $^{46}$Fe, the 2p daughter of $^{48}$Ni. These four 
decay events, one of possible 2p origin and three from $\beta$-delayed decays, were also used to determine the 
half-life of $^{48}$Ni (T$_{1/2}$= 2.1$^{+2.1}_{-0.7}$ms). Although this single possible 2p decay event is not 
considered sufficient to claim the observation of 2p radioactivity for $^{48}$Ni,  these data were nonetheless 
used to determine the 2p branching ratio to be 25$^{+29}_{-19}$\% and the  partial 2p half-life to be 
T$_{1/2}^{2p}$=8.4$^{+12.8}_{-~7.0}$ms~\cite{dossat05}. 

\subsection{Direct observation of two protons in ground-state decay of $^{45}${\rm Fe}}

 From all 2p emitters observed up to now, the ground-state emitter $^{45}$Fe is the best 
candidate to search for a clear signal of correlated emission and to study in 
detail the decay mechanism. 

As just mentioned, 2p emission could only be
governed by phase space or there could be more or less strong angular and energy correlations. In order to 
study these questions, new experimental setups had to be imagined, the basic limitation of the
silicon-detector setups being that the 2p emitters are deeply implanted in a silicon detector, which then
does not allow the detection of individual protons, but rather of the total decay energy and the half-life
only.

\begin{figure}[hht]
\begin{center}
\resizebox{.4\textwidth}{!}{\includegraphics[angle=-0]{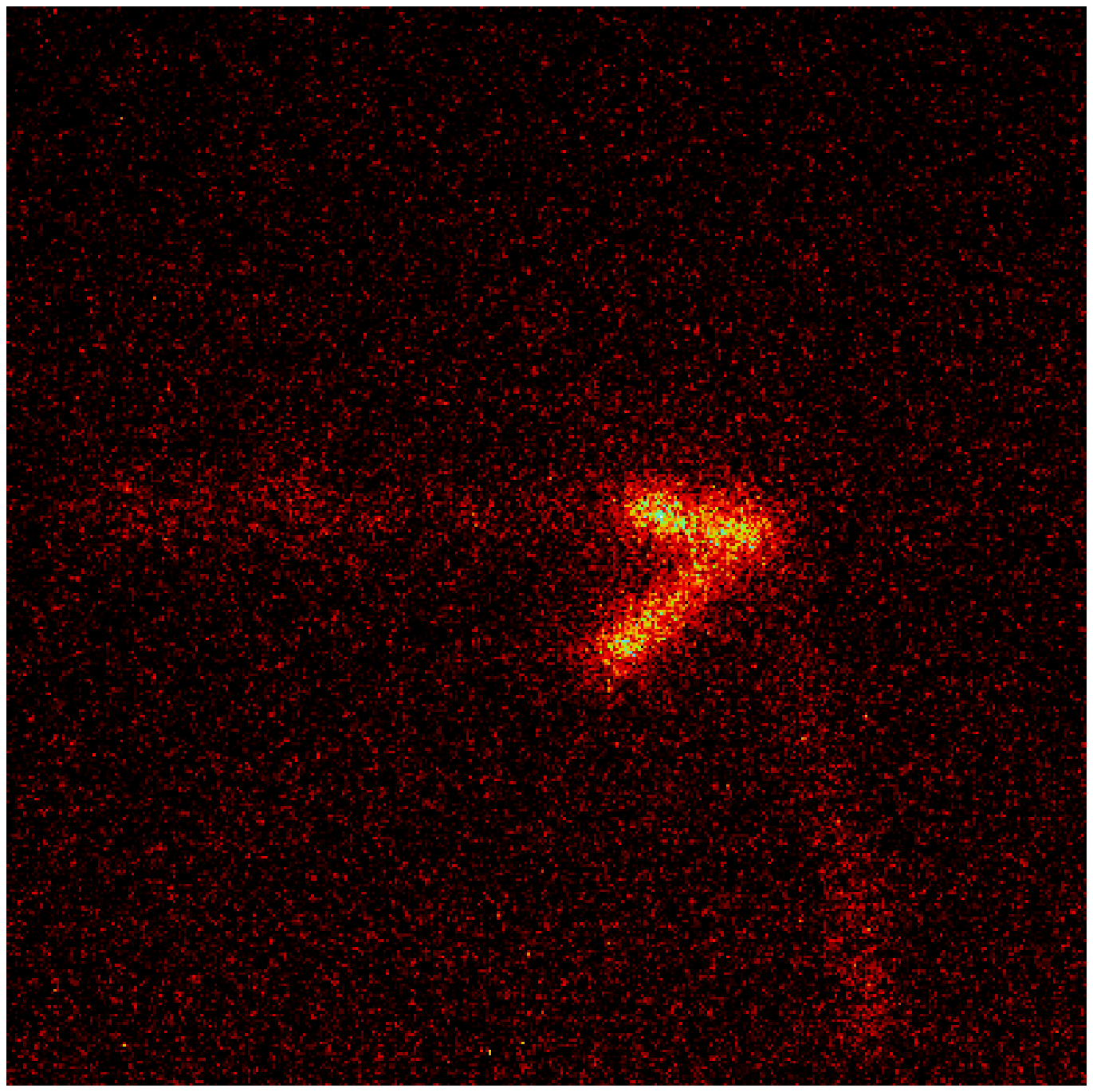} }
\resizebox{.58\textwidth}{!}{\includegraphics[angle=-0]{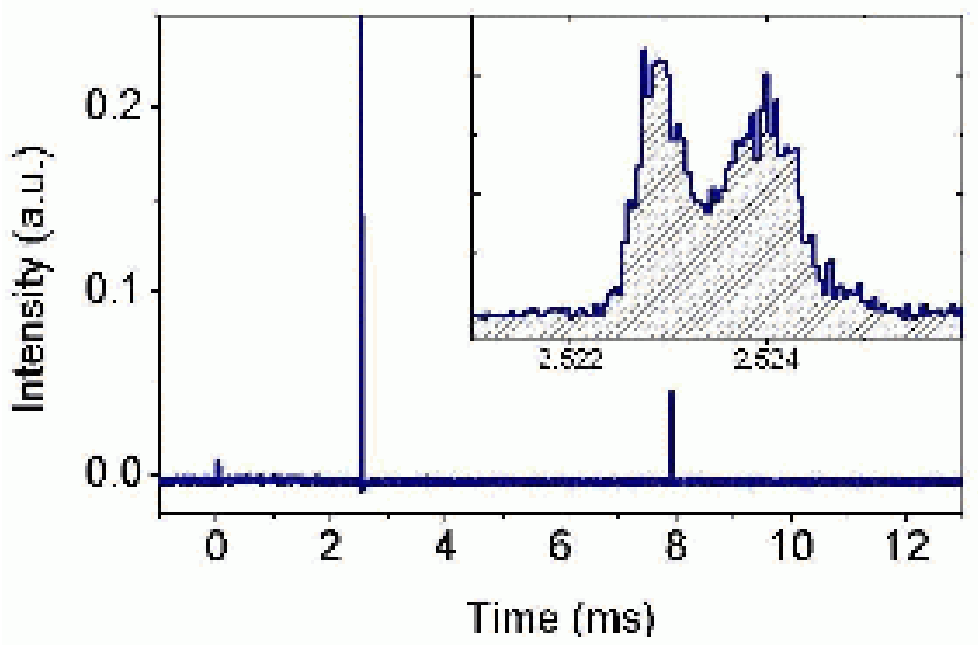} }
\end{center}
\caption{An example of a two-proton decay event of $^{45}$Fe recorded by the  Warsaw TPC detector.  
         Left: an image taken by the CCD camera in a 25 ms exposure. The faint track of a $^{45}$Fe ion entering the chamber 
         from the left is seen. The two bright, short tracks are low-energy protons emitted approximately 2.5~ms after 
         the implantation. A weaker and longer track, going from the decay vertex to the lower right, represents a 
         high-energy proton emitted about 8~ms after the implantation which is due the decay of $^{43}$Cr, the 2p decay daughter 
         of $^{45}$Fe. Right: the time profile of the total light intensity, measured by the PMT, provides the time sequence 
         of the decay events. In the inset, a magnified part is plotted showing in detail the 2p emission. Figure courtesy 
         of M. Pf{\"u}tzner.}
\label{fig:2pmsu}
\end{figure}

The solution to this problem is the use of gas detectors capable to visualize the traces of the emitted protons.
Therefore, time projection chambers (TPC) were set up in Bordeaux~\cite{blank07tpc} and Warsaw~\cite{miernik07}. 
The Bordeaux detector is based on gas electron multiplier technology~\cite{sauli97} and a double-sided microgroove 
detector~\cite{bellazzini99} 
with an application-specific integrated-circuit  read-out, whereas the Warsaw detector uses an optical read-out system 
by means of a digital camera and a photomultiplier tube (PMT) which registers the time sequence. 
Both TPC detectors have recently taken first data~\cite{giovinazzo07,miernik07b} (see Figs.~\ref{fig:2pevent},~\ref{fig:2pmsu}). 
Due to the highest production rates, $^{45}$Fe was the first 2p emitter to be studied with these devices.

Due to the higher statistics, the data taken at Michigan State University with the Warsaw TPC allowed to compare the experimental 
data to predicitons of the ${\rm S}\ell^2{\rm M}$ model of Grigorenko et al. model~\cite{grigorenko01}, for which nice agreement 
for the angular and energy 
distribution was reached~\cite{miernik07b} indicating e.g. an equal sharing of the energy between the two protons.

\section{Conclusion for the experimental part}

The observation of ground-state 2p radioactivity for $^{45}$Fe, $^{54}$Zn and possibly for
$^{48}$Ni allowed this type of radioactivity to be established as a new nuclear decay mode. Although only based on theoretical 
mass predictions, these decays are believed to be simultaneous emissions of the two protons. A sequential decay could have a 
sizable contribution only 
if the $Q_{1p}$ values were off by a few hundred keV what is
rather unlikely. Nonetheless, the assumption that the decay is indeed simultaneous still awaits 
experimental verification though, as will be discussed in Sect. 9,  a symmetric energy distribution of individual protons 
emitted in the 2p decay of $^{45}$Fe~\cite{miernik07b} is a strong indication for a predominance of the simultaneous 2p 
emission. Table~\ref{tab:2p-emitters} summarizes the experimental information available for $^{45}$Fe, $^{48}$Ni, and $^{54}$Zn.

\renewcommand{\arraystretch}{.95}
\begin{table}[hht]
\begin{center}
\begin{tabular}{|c|c|c|c||c|}
\hline  \rule{0pt}{1.3em}
                                  &~2p decay energy ~     &~half-life~           & ~branching ratio~       &~partial half-life~\\ 
                                  & (MeV)                 &  (ms)                &                         &  (ms) \\ 
[0.5em] \hline \rule{0pt}{1.3em}
$^{45}$Fe                         &   1.151$\pm$0.015     &2.5$^{+0.2}_{-0.2}$   &   0.65$\pm$0.05         &   3.9$^{+0.4}_{-0.4}$  \\ 
[0.5em] \hline \rule{0pt}{1.3em}
$^{48}$Ni                         &   1.35$\pm$0.02       &2.1$^{+2.1}_{-0.7}$   &   0.25$^{+0.29}_{-0.19}$&   8.4$^{+12.8}_{-7.0}$  \\ 
[0.5em] \hline \rule{0pt}{1.3em}
$^{54}$Zn                         &   1.48$\pm$0.02       &3.2$^{+1.8}_{-0.8}$   &   0.87$^{+0.10}_{-0.17}$&   3.7$^{+2.2}_{-1.0}$  \\ 
[0.5em] \hline 
\end{tabular}
\end{center}
\caption[]{Decay energies, half-lives, branching ratios, and partial half-lives as determined in three 
           independent experiments for the decay of $^{45}$Fe~\cite{giovinazzo02,pfuetzner02,dossat05,miernik07b} as well as 
           for $^{48}$Ni~\cite{dossat05}, and $^{54}$Zn~\cite{blank05zn54}. 
           }
\label{tab:2p-emitters}
\end{table}

To get a coherent picture of this new decay mode, more 2p emitters have to be studied experimentally.
However, as these nuclei are, by definition, situated beyond the proton drip line, their production is
rather difficult and the rates are very low. Rather high rates can be achieved today at the SISSI/LISE3 
facility of GANIL with about 10-15 $^{45}$Fe per day, about 2 $^{54}$Zn per day, and roughly 
1 $^{48}$Ni per day under optimal conditions. It was recently shown that at the coupled-cyclotron facility and the 
A1900  separator of the NSCL even higher rates can be achieved. Other candidates like $^{59}$Ge, 
$^{63}$Se or $^{69}$Kr might be reached in the near future at the new RIKEN Radioactive Ion Beam Facility (RIBF) in Tokyo, Japan.

\section{Theoretical description of two-proton radioactivity}

The nuclear many-body Hamiltonian is supposed to describe all nuclei that can exist and not merely one nucleus of 
a given number of protons and neutrons. In this sense, a nucleus is never a closed, isolated quantum system but 
communicates with other nuclei through virtual excitations, decays and captures. The communication is broken and 
the nucleus becomes artificially closed if the subspace of continuum states is excluded in the network of coupled 
systems. Obviously, the closed quantum system idealization of  a real many-body system has very different features 
from those observed~\cite{okolowicz03,dobaczewski98}, in particular in the neighborhood of each reaction 
threshold~\cite{dobaczewski07,baz69}. 

Most of the nuclear states are embedded in the continuum of decay channels due to which they get a finite lifetime. 
The initial impulse for a mathematical formulation of the shell model (SM) for open quantum systems goes back to 
Feshbach and Fano~\cite{feshbach58,feshbach62,fano61} who introduced the two subspaces of Hilbert space: (i) the 
subspace of discrete states, and (ii) the subspace of real-energy scattering states, respectively. Feshbach 
achieved a unified description of direct processes at the short-time scale and of compound nucleus processes at the long-time 
scale~\cite{feshbach58,feshbach62}. A unified  description of both nuclear structure and nuclear reactions in a 
single theoretical framework became possible only in the continuum 
SM~\cite{mahaux69,philpott77,halderson80,barz77,barz78,rotter91}. 

First realistic studies using the continuum SM have been presented only at the turn of the century in the 
framework of SMEC~\cite{bennaceur99,bennaceur00} (see also Ref.~\cite{okolowicz03} for a recent review). 
This model is based on the completeness of the single-particle basis 
consisting of bound orbits and a real-energy  scattering continuum.   Single-particle resonances are regularized, i.e. 
they are included in a discrete part of the spectrum after removing scattering tails which are incorporated 
in the embedding continuum.  Subsequently, the new set of discrete states and the non-resonant scattering 
states are reorthogonalized. Configuration mixing in the valence space (the internal mixing) is given by an 
effective SM interaction. Coupling of discrete states to the continuum of decay channels is calculated 
microscopically and leads to the so-called external mixing of SM wave functions. The decay channels are 
obtained in the $S$-matrix (scattering-matrix) formalism (see Ref.~\cite{okolowicz03} and references quoted therein). 

Another approach referring to the continuum SM has been proposed recently in 
Refs.~\cite{volya03,volya05}. Internal mixing in this approach is given by an effective SM  Hamiltonian but 
an external mixing is neglected. This scheme has been applied to the description 
of observables in the one- and two-particle continuum~\cite{volya05}.

The genuine three-body decay models for 2p radioactivity have been proposed in Refs. 
\cite{federov94,nielsen01,grigorenko03d,grigorenko03c} using the hyperspherical harmonics method. These models 
combine the cluster model description of the decaying system with a treatment of three-body asymptotic states of 
charged particles. Rigorous treatment of a three-body kinematics in these models allows to study both angular 
and energy/momentum correlations between decaying fragments. 

A different approach for describing many-body open quantum systems is provided by a generalization of SM in 
the complex-momentum plane (the GSM)~\cite{michel02,michel04,idbetan02}. 
This general formalism with no restriction on the number of particles 
in the scattering continuum has not yet been applied for a description of experimentally observed 2p decays.

\subsection{Shell Model Embedded in the Continuum with one particle in the scattering continuum \label{SMEC_1p}}

A general theory of the decaying many-body system is provided by the open quantum system formulation of the SM. 
Before we turn to the description of 2p decay, let us introduce the basic features of SMEC  in the simpler case of the 
one-particle continuum. In SMEC, the function space consists of two sets: 
the space of ${\cal L}^2$-functions, as in the nuclear 
SM, and the space of scattering states (decay channels) embedding SM 
states. The decay is a result of the coupling between discrete states and decay channels. These two sets 
of wave functions are found by solving the Schr\"odinger equation for discrete states:
\begin{eqnarray}
\label{SMeq1}
H_{SM}|\Phi_i\rangle=E_i^{(SM)}|\Phi_i\rangle ~ \ ,
\end{eqnarray}
where $H_{SM}$ is the SM Hamiltonian, and for scattering states:
\begin{eqnarray}
\label{SMeq2}
\sum_{c'}(E-H_{cc'})\xi_E^{c'(+)}=0 ~ \ ,
\end{eqnarray}
where $H_{cc}=H_0+V_{cc}$ in (\ref{SMeq2}) is the standard coupled-channel Hamiltonian. Different channels 
are defined by coupling the motion of an unbound particle relative to the residual nucleus with $(A-1)$ 
particles in a discrete SM state $|\Phi_j^{A-1}\rangle$. 
$\xi_E^{c(+)}$ are scattering states projected on the channel $c$, and the channel numbers $c$ are defined 
by the quantum numbers of of the states $j$ of the residual nucleus and those of the unbound particle which 
are coupled to the total quantum numbers $J^{\pi}$ and $T$ of the system.

Using these two function sets: ${\cal Q}\equiv \{\Phi_i^{A}\}$, ${\cal P}\equiv \{\xi_E\}$, 
one defines projection operators:
\begin{eqnarray}
{\hat Q} = \sum_{i=1}^N|\Phi_i^A\rangle\langle\Phi_i^A| ~~~~; ~~~~~
{\hat P} = \int_{0}^\infty dE|\xi_E\rangle\langle\xi_E|
\end{eqnarray}
satisfying
${\hat Q}|\xi_E\rangle = 0$, ${\hat P}|\Phi_i^A\rangle = 0$,
and projected Hamiltonians: ${\hat Q}H{\hat Q}\equiv H_{QQ}$ and ${\hat P}H{\hat P}\equiv H_{PP}$. 
The closed system Hamiltonian $H_{QQ}$ is the nuclear SM Hamiltonian 
$H_{SM}$, whereas $H_{PP}$ is given by  $H_{cc}$. 
To construct a complete solution in ${\cal Q}+{\cal P}=I_d$:
$$|\Psi_E\rangle={\hat Q}|\Psi_E\rangle+{\hat P}|\Psi_E\rangle ~ \ ,$$ 
one has to find a wave function connecting ${\cal Q}$ and ${\cal P}$ subspaces. This wave function 
$\omega_i^{(+)}$ is obtained as
a solution of non-homogeneous coupled-channel equations: 
\begin{eqnarray}
\label{SMeq5}
|\omega_i^{(+)}(E)\rangle=G_P^{(+)}(E)|w_i\rangle~ \ ,
\end{eqnarray} 
where  {\bf{$|w_i\rangle=H_{PQ}|\Phi_i\rangle$}} is the source term, 
$$G_P^{(+)}(E)=\lim_{\epsilon \to 0}[E+i\epsilon-H_{PP}]^{-1}$$ 
 is the Green's function in ${\cal P}$ subspace, $E$ is the total energy of the nucleus $[A]$, and 
$H_{PQ}\equiv {\hat P}H{\hat Q}$. Using the three function sets:  $\{\Phi_i^{A}\}$, 
$\{\xi_E\}$, and $\omega_i^{(+)}(E)$, one can find the solution in the total function space:
\begin{equation}
\hspace*{-1.5cm}
|\Psi_E\rangle=|\xi_E\rangle 
+\sum_{i,k}(|\Phi_i^A\rangle+|\omega_i^{(+)}(E)\rangle)\langle\Phi_i^A|
(E-{\cal H}_{QQ}(E))^{-1}|\Phi_k^A\rangle\langle\Phi_k^A|H_{QP}|\xi_E\rangle ~ \ .
\end{equation}
${\cal H}_{QQ}(E)$ is the energy-dependent effective Hamiltonian in the function space of discrete states:
\begin{eqnarray}
\label{SMeq3}
{\cal H}_{QQ}(E)=H_{QQ}+H_{QP}G_P^{(+)}(E)H_{PQ}
\end{eqnarray}
which takes into account couplings to decay channels. In that sense, ${\cal H}_{QQ}$ is the open 
quantum system Hamiltonian in ${\cal Q}$ subspace. 

The second term on the right-hand side of Eq. (\ref{SMeq3}) provides an imaginary part of energy 
eigenvalues for states above the particle emission threshold. The real part of this term contributes 
to the real part of energy eigenvalues both below and above the threshold. For that reason, it would 
be more appropriate in phenomenological applications to fit matrix elements of ${\cal H}_{QQ}$ and 
not of $H_{QQ}$ to nuclear spectra in a given mass region. This ambition task has not yet been accomplished.

 Energies $E_i$ and widths $\Gamma_i$ of the resonance states derive from the solutions 
of fixed-point equations:
\begin{eqnarray}
\label{fixed_point}
E_i = {\tilde E}_i(E=E_i) ~~~~; ~~~~
\Gamma_i = {\tilde \Gamma}_i(E=E_i)
\end{eqnarray}
where functions ${\tilde E}_i(E)$ and ${\tilde \Gamma}_i(E)$ follow from (complex) eigenvalues of 
${\cal H}_{QQ}$. 

The eigenvalues of ${\cal H}_{QQ}$ have a physical meaning if the total function space is divided into the 
subspaces of the {\em system} (the subspace ${\cal Q}$) and the {\em environment} (the subspace ${\cal P}$) 
as follows: the system contains all resonance-like phenomena while the environment describes the smooth part 
in the energy region considered. This operational definition of subspaces implies that in SMEC one 
has to construct carefully the single-particle basis. The narrow single-particle resonances in the continuum are 
included into the set  of discrete states either as resonance anamneses \cite{faes07} or as quasi-bound 
single-particle states \cite{wang70}. The new definition of discrete single-particle states implies a 
redefinition of the continuous spectrum, as described in Ref. \cite{okolowicz03}.

\subsection{Theory of two-proton radioactivity}

The description of  Borromean (halo) systems and two-nucleon decays involves couplings to the two-nucleon 
decay channels. In the 2p decay, since two protons do not form a bound two-particle system, one neither 
can uniquely specify the final 
state of two protons nor the transition path leading to this state. Various transition paths may interfere
making distinction between different decay mechanisms difficult if not impossible.
In this respect, the 2p decay is qualitatively different from any standard two-body 
decays such as the proton-, neutron-  or $\alpha$-particle decays. 

A generalization of the SMEC formalism for larger number of particles in the scattering 
continuum is conceptually similar to the one-particle case discussed above (cf Sect. \ref{SMEC_1p}). 
For the description of two-particle continuum, one has to introduce  a subspace ${\cal  T}$ with 
two particles in the continuum, and ${\hat T}$ the corresponding projection operator.  The completeness 
relation in total function space is then: ${\cal Q}+{\cal P}+{\cal T}=1$. 
Similarly as in Sect.  \ref{SMEC_1p}, one can decompose the Hamiltonian $H$ into components 
acting in different subspaces and the coupling terms between those different subspaces. The ${\cal Q}$ 
subspace effective Hamiltonian ${\cal H}_{QQ}(E)$ 
can be written as~\cite{rotureau06}: 
\begin{eqnarray}
{\cal H}_{QQ}(E)&=&H_{QQ}+H_{QP}G_{P}^{+}(E)H_{PQ} \nonumber \\
&+&[H_{QT}+H_{QP}G_{P}^{+}(E)H_{PT}]\tilde{G}_{T}^{+}(E)[H_{TQ}+H_{TP}G_{P}^{+}(E)H_{PQ}] 
\label{H_eff_T}
\end{eqnarray}
The first term on the right-hand side is the closed quantum system Hamiltonian, the 
second term describes a coupling with the one-particle continuum, and the third term corresponds to all 
possible couplings with the two-particle continuum. $\tilde{G}_{T}^{+}(E)$ is the Green function in 
${\cal T}$
$$\tilde{G}_{T}^{+}(E)=\lim_{\epsilon \to 0}[E+i\epsilon-H_{TT}-H_{TP}G_{P}^{+}(E)H_{PT}]^{-1}$$ modified 
by couplings to ${\cal P}$. Similarly as in the case of one-particle continuum, one defines the source term
\begin{eqnarray}
\label{intro_01}
|w_i\rangle=[H_{TQ}+H_{TP}G_{P}^{(+)}(E)H_{PQ}]|\Phi_i\rangle  \label{source_T} 
\end{eqnarray}
and the continuation $|\omega^{+}_i\rangle$ of SM wave function $|\Phi_i\rangle$ into the two-particle 
continuum:
\begin{eqnarray}
\label{intro_02}
|\omega^{+}_i\rangle=\tilde{G}_{T}^{(+)}(E)[H_{TQ}+H_{TP}G_{P}^{(+)}(E)H_{PQ}]|\Phi_i\rangle 
=\tilde{G}_{T}^{(+)}(E)|w_i\rangle ~ \ .
\label{omega_T}
\end{eqnarray}
Matrix elements  of the operator responsible for ${\cal Q} - {\cal T}$ couplings 
(the third term on the right-hand side of Eq.  (\ref{H_eff_T})) can be expressed as 
the overlap between the source term and the function $\omega^{+}_j$ (see Eqs.  (\ref{intro_01}) and 
(\ref{intro_02})).  

Effective Hamiltonian ${\cal H}_{QQ}(E)$ can be written in various equivalent forms.  For our purpose, 
it is convenient to extract the direct coupling term between ${\cal Q}$ and ${\cal T}$ subspaces:
\begin{eqnarray}
&&{\cal H}_{QQ}(E)=H_{QQ}+H_{QT}G_{T}^{(+)}(E)H_{TQ} \nonumber \\
&+&[H_{QP}+H_{QT}G_{T}^{(+)}(E)H_{TP}]\tilde{G}_{P}^{(+)}(E)[H_{PQ}+H_{PT}G_{T}^{(+)}(E)H_{TQ}] 
~ \ .  
\label{H_eff_T2}
\end{eqnarray}
$\tilde{G}_{P}^{(+)}(E)$ in (\ref{H_eff_T2}) stands for ${\cal P}$-subspace Green function modified by 
couplings to ${\cal T}$
$$\tilde{G}_{P}^{(+)}(E)=\lim_{\epsilon \to 0}[E+i\epsilon-H_{PP}-H_{PT}G_{T}^{(+)}(E)H_{TP}]^{-1} ~ \ ,$$ where
$$G^{(+)}_{T}(E) = \lim_{\epsilon \to 0}\left[ E+i\epsilon-H_{TT}\right]^{-1}$$ is the Green function 
in ${\cal T}$. 

${\cal H}_{QQ}(E)$ (cf Eqs. (\ref{H_eff_T}) and (\ref{H_eff_T2})) describes all 
emission processes involving one and two nucleons (two protons). Observed 2p partial decay width is a sum of 
contributions of different interfering processes which cannot be disentangled experimentally. In theoretical 
studies, one can always isolate a given processes by switching-off all others and calculate the contribution 
of such an isolated process to the 2p partial width. However, relative weights of different isolated processes 
do not include interference effects and, hence, the resulting picture of the 2p decay is somewhat distorted. In this 
sense, and contrary to the two-body decays, the interpretation of experimental 2p decay data is unavoidably model dependent.

In the observed 2p decays, certain emission scenarios may 
be less probable, so it is instructive and certainly less cumbersome to consider limiting cases of the 2p 
decay. For example, if ${\cal P}\leftrightarrow {\cal T}$  couplings are weak then  the dominant process is 
a direct 2p emission described by 
\begin{eqnarray}
{\cal H}_{QQ}^{(dir)}(E)=H_{QQ}+H_{QP}G_{P}^{(+)}(E)H_{QP}+H_{QT}G^{(+)}_{T}(E)H_{TQ} ~ \ ,
 \label{H_eff_direct}
\end{eqnarray}
where the second term on the right-hand side is  responsible for couplings to one-nucleon decay channels. On the contrary, if
an intermediate system  $[A-1]$ plays an important role, then ${\cal Q}\leftrightarrow {\cal T}$ couplings  
can be neglected  and the effective Hamiltonian reads:
\begin{eqnarray}
{\cal H}_{QQ}^{(seq)}(E)=H_{QQ}+H_{QP}\tilde{G}_{P}^{(+)}(E)H_{PQ} ~ \ .
\label{H_eff_seq2}   
\end{eqnarray}
This operator describes a standard sequential 2p emission via an intermediate resonance in the 
$[A-1]$ nucleus. 

The sequential 2p emission process may also occur via continuum states which are correlated by 
either weakly bound or by weakly unbound states of  $[A-1]$ nucleus. This process which goes 
through the tail of an energetically inaccessible state, is called the {\em virtual sequential} 
two-body decay~\cite{garrido05}. For its study, it is important to separate 1p- and 2p-decays in the effective Hamiltonian:
\begin{eqnarray}
{\cal H}_{QQ}^{(seq)}(E)&=&H_{QQ}+H_{QP}G_{P}^{(+)}(E)H_{PQ} \nonumber \\
&+&H_{QP}\tilde{G}_P^{(+)}H_{PT}G_{T}^{(+)}(E)H_{TP}G_{P}^{(+)}(E)H_{PQ} ~ \ .
\label{3_express}
\end{eqnarray}
The second term in Eq. (\ref{3_express})  describes the 1p decay, whereas the third term describes 
the genuine sequential decay.

\subsubsection{Virtual sequential two-proton emission \label{sub_seq}} 

The sequential 2p decay via the correlated scattering continuum of $[A-1]$ nucleus  (the virtual sequential 
emission process) is an important component of the 2p decay mechanism.
An effective Hamiltonian for this process ( cf Eq. (\ref{3_express})) describes interference effects between 
the 1p decay and the sequential 2p decay, both for closed (virtual ${\cal Q}$-${\cal P}$ excitations) and opened 
(1p decays) 1p decay channels. Diagonalizing ${\cal H}_{QQ}^{(seq)}(E)$, one obtains energies  of states in 
$[A]$ nucleus as well as their widths associated with the emission of one and two protons. In general, these 
two decay modes cannot be separated one from another. However, since couplings responsible for the 2p decay are  smaller 
than those associated with the 1p decay, one may proceed in two steps. In the first step, one neglects the term
$H_{QP}\tilde{G}_P^{(+)}H_{PT}G_{T}^{(+)}(E)H_{TP}G_{P}^{(+)}(E)H_{PQ}$ (cf Eq. (\ref{3_express})) responsible for the 2p decay, and
diagonalizes the remaining operator  in a SM basis $\{|\Phi^{A}\rangle\}$. This provides new basis 
states $\{{|\tilde \Phi}^{A}\rangle\}$ which are linear combinations of SM states. In the second step, using 
those mixed SM states  one calculates the sequential 2p emission width of 
a decaying state $|{\tilde \Phi}^{A}_{i}\rangle$, i.e. one calculates an imaginary part of the matrix element 
$\langle {\tilde \Phi}^{A}_{i}|H_{QP}\tilde{G}_P
H_{PT}G_{T}^{(+)}(E)H_{TP}G_{P}^{(+)}(E) H_{PQ}|{\tilde \Phi}^{A}_{i}\rangle ~.$ 

Let us assume that different steps of the sequential process are uncorrelated, i.e. the 
first emitted proton is a spectator of the second emission. This implies: 
\begin{eqnarray}
H_{PP}& \rightarrow& H_{Q'Q'}+{\rm {\hat p}}h_0{\rm {\hat p}} \nonumber \\
H_{TT} &\rightarrow& H_{P'P'}+{\rm {\hat p}}h_0{\rm {\hat p}} \nonumber
\end{eqnarray}
where primed quantities refer to $(A-1)$-nucleon space. In the ${\cal Q}{'}$ subspace, $(A-1)$ 
nucleons are in quasi-bound 
single-particle orbits, whereas in the ${\cal P}{'}$ subspace, one nucleon is in the continuum 
and $(A-2)$ nucleons are in quasi-bound 
single-particle  orbits. $h_0$ is a one-body potential describing an average effect 
of $(A-1)$-particles on the emitted proton and ${\rm {\hat p}}$ denotes a projector on one-particle 
continuum states. With this identification, $H_{PT}$ becomes a coupling between newly defined 
${\cal Q}{'}$ and ${\cal P}{'}$ subspaces. 

The virtual sequential 2p emission appears always if $Q_{2p}>0$. In that sense, the virtual sequential process is a 
component of any quantum-mechanical three-body decay. In the approximate scheme employed in SMEC, the virtual sequential 
2p decay process competes with the 
diproton mode  even for closed 1p emission channels.

\subsubsection{Diproton emission  \label{cluster_theo}}

The effective Hamiltonian  (\ref{H_eff_direct}) describes both the three-body direct 2p decay 
and its limit of two sequential two-body decays. In this section, we shall discuss the 
latter two-step scenario~\cite{brown03}. In the first step, protons are emitted 
as a '$^2$He' -cluster which, in the second step, disintegrates outside of a nuclear potential  
due to the final state interaction~\cite{watson52,migdal53}. The final state 
proton-proton interaction is taken into account by the density $\rho(U)$ of proton states corresponding 
to the intrinsic energy $U$ in the proton-proton system~\cite{barker01}. 
The total system ${\rm [A-2]}\otimes{\rm [2p]}$ is described as the two-body system in a mean-field  
$U_{0}$:
\begin{eqnarray}
H_{TT}= {\hat T}^{(cl)} \left [ {\tilde H}^{(A-2)}+ {\tilde H}^{(cl)}+ \frac{P^{2}}{2\mu} + U_{0} \right ] {\hat T}^{(cl)} ~ \ ,
\label{H_tt_cluster}
\end{eqnarray}
where ${\tilde H}^{(A-2)}$ is the intrinsic Hamiltonian of the daughter nucleus  $[A-2]$, and  
${\tilde H}^{(cl)}$ is the intrinsic Hamiltonian of '$^2$He'-cluster. The cluster is described 
as a particle of mass $M=2M_{p}$ ($M_{p}$ denotes the proton mass) and charge $Z=2$. $P^{2}/2\mu$ 
is the intrinsic kinetic energy of the system ${\rm [A-2]}\otimes{\rm [2p]}$, and $\mu$  is the 
reduced mass of the system.  ${\hat T}^{(cl)}$ is the projection operator on the subspace of cluster 
states in the continuum of the potential $P^{2}/2\mu + U_{0}$. 
The 2p decay width is given by the matrix element:
\begin{eqnarray}
\label{eq4}
\Gamma_{[2p]}&=&-2 Im\left( \langle{\tilde \phi}^{(int)}_{i}|H_{QT}G_{T}^{(+)}(E)H_{TQ}|{\tilde \phi}^{(int)}_{i}\right) 
\rangle \\ \nonumber &=& -2 Im \left(\langle w_i^T|\omega_i^{T,(+)}\rangle  \right) ~ \ ,
\end{eqnarray}
where $|{\tilde \phi}^{(int)}_{i}\rangle$ is the intrinsic state corresponding to a mixed state 
$|{\tilde\Phi}_{i}^{A}\rangle$ in the parent nucleus, $|w_i^T\rangle = H_{TQ}|{\tilde \phi}^{(int)}_{i}\rangle$ 
is the source term,  and $|\omega_i^{T,(+)}\rangle=G_{T}^{(+)}(E)H_{TQ}|{\tilde \phi}^{(int)}_{i}\rangle$ 
is a continuation of $|{\tilde \phi}^{(int)}_{i}\rangle$ in ${\cal T}$. The Coulomb interaction is included 
as an average field and does not enter in $H_{TQ}$. 
The matrix element (\ref{eq4}) is the integral over an intrinsic energy $U$ of the overlap of a 
channel-projected source term and a channel-projected continuation  in ${\cal T}$ of an intrinsic state  
$|{\tilde \phi}^{(int)}_{i}\rangle$, weighted by the proton state density  
$\rho(U)$~\cite{rotureau05,rotureau06}. Working with intrinsic states
allows to account for the recoil correction.

The decay energy is divided into an intrinsic energy of the '$^2$He'-cluster and its 
center of mass motion energy according to the proton-state density~\cite{rotureau05,rotureau06}. 
In this way,  intrinsic degrees of freedom of the two protons are reduced and, therefore,  the three-body 
decay is reduced to the two sequential two-body decays.

\paragraph{${\rm R}$-matrix model of Brown and Barker  \label{Barker_theo}}

This model extends the R-matrix approach for SM description of a three-body decay~\cite{barker01,brown03}.  
Similarly as in the SM approach for two-body decays,
one may assume that the three-body decay is independent of the small-distance many-body dynamics. This opens 
for a definition of three-body spectroscopic factors which can be computed in the SM framework~\cite{brown03} 
in an analogous way as the preformation factors used to describe the
$\alpha$-particle decay. 

The R-matrix model of a diproton decay~\cite{barker01,brown03} includes the $s$-wave 
proton-proton interaction as an intermediate state. This final state interaction is crucial for a quantitative 
description of experimental data, changing the diproton lifetimes by several order of magnitudes. 
$^{2}$He preformation factor is taken into account by $^{2}$He spectroscopic factors $S_{[2p]}$. To calculate them, 
the SM wave function is projected onto the $0s$ internal relative wave function for a '$^{2}$He'-cluster~\cite{brown03}: 
$$S_{[2p]}=\left( \frac{A}{A-2} \right)^{\lambda}G^2C(A,Z)~.$$
In $fp$ shell model sace, $G^2=5/16$, $\lambda=6$. $C(A,Z)=|\langle\Psi(A-2.Z-2)|\Psi_c|\Psi(A,Z)\rangle|^2$ is 
the overlap for the $^{2}$He wave function $\Psi_c$ with $L=0$, $S=0$, and $T=1$ in the $SU(3)$ basis. 

The extended R-matrix model ~\cite{brown03} neglects both the small-distance dynamics 
and the virtual scattering to continuum states. 
The latter process changes the $^2$He emission probability by modifying the wave function 
of a decaying state and changing the $^2$He preformation factor~\cite{rotureau05}.
This change is expected to be particularly strong for states close to the 2p
emission threshold(s), e.g. the ground states of $^{45}$Fe, 
$^{48}$Ni and $^{54}$Zn. Reasons for the continuum-induced change of $^2$He preformation probability or one-nucleon 
spectroscopic factors~\cite{chatterjee06,michel07a,michel07b} 
 are the same as for a dominance of cluster states close to their respective decay thresholds~\cite{ikeda68}:  both 
 phenomena are consequences of the 'alignment' of near-threshold states with the channel 
 state~\cite{okolowicz03,luo02,michel05,okolowicz06,dobaczewski07}.

It is not easy to assess quantitative consequences of those simplifications in the extended R-matrix model. In the 
following sections, we shall try to address
 this problem by (i) comparing SMEC results for diproton  decays with those of an extended R-matrix approach, and 
 (ii) by analyzing SMEC results for virtual sequential 2p emission with and without external mixing.

\subsubsection{Direct 2p emission with three-body asymptotics \label{direct_theo}}

In general, structures in the continuum are more difficult to treat than bound-state problems.
Firstly, the three-body problem for unbound systems is much less established, although studied for 
short and long-range interactions~\cite{glockle96,tanner00,michel04}.
Secondly, the Coulomb problem with three-body asymptotics is still considered unsolved~\cite{lin95,rescigno99,alt04} 
even though  an important progress has been reported recently in the description of proton-deutron scattering and of 
three-nucleon electromagnetic reactions involving $^3$He~\cite{deltuva05}. 
An infinite range of the Coulomb interaction in channel-coupling potentials does not allow to decouple equations  at  
infinity. Consequently, an asymptotic behaviour of 
$\omega^{(+)}_{j,c}(\rho)$ cannot be defined without a certain approximation. One approximate way to proceed is to 
neglect off-diagonal channel-coupling potentials above a certain value of the hyperradius  and to define an effective 
Sommerfeld parameter from diagonal coupled-channel 
potentials~\cite{grigorenko03d}. This technique has been applied in the three-cluster model of the 2p 
emission~\cite{grijo01,grigorenko03d,grijomu00}.

In order to calculate the direct 2p emission probability, one has to evaluate 
the matrix element (\ref{eq4}) 
without a simplifying assumption of two sequential two-body decays. Even though this three-body problem has been 
conveniently formulated in Jacobi coordinates~\cite{rotureau06}, 
its numerical solution in SMEC has not yet been given. Firstly, if the residual interaction is a 
contact force, then the interaction of two protons 
in the continuum leads to the ultra-violet divergence. In principle, this divergence could be avoided 
using a finite-range residual interaction. However, such an interaction leads to non-local channel-coupling 
potentials and, hence, equations for channel-projected $\omega_{j,c}^{(+)}$ functions become 
integro-differential coupled-channel equations. 

\paragraph{The three-body cluster model: the Hyperspherical Adiabatic Expansion method \label{garr_theor}}

In this approach, the dynamics of $[A]$-particle system is reduced to a three-body dynamics which 
is described by solving Faddeev equations~\cite{federov94,nielsen01}.  
The three-body wave function is a sum of three Faddeev components, each of them expressed in one of 
the three possible sets of Jacobi coordinates $\{{\bf x}_i, {\bf y}_i\}$. Each component is then 
expanded in terms of a complete set of hyperangular functions. 
Essential quantities in this approach are the effective radial potentials in the  hyperradius variable:
\begin{eqnarray}
\rho^2\equiv\frac{1}{mM}\sum_{i<k}m_im_k{\bf r}_{ik}^2=({\bf x}_j^2+{\bf y}_j^2)
\end{eqnarray}
with ${\bf r}_{ik}^2=({\bf r}_i-{\bf r}_k)^2$, where ${\bf r}_i$ is the coordinate of particle $i$, 
$M=\sum_{l=1}^3m_l$ and $m$ is an arbitrary normalization mass. The set of indices $(i,j,k)$ is a 
permutation of $(1,2,3)$. ${\bf x}_j$, ${\bf y}_j$ are proportional to the distance between two 
particles and the distance between their center of mass and the third particle, respectively. 

The behaviour of effective radial potentials (the hyperspherical adiabatic potentials) as a function 
of $\rho$ for different angular momentum quantum numbers determines the structure of a three-body 
system at both small and large distances. For short-range interactions, the lowest hyperspherical 
adiabatic potential usually dominates in the expansion of the wave function. If both short and 
long-range interactions are present, the wave-function expansion at large distances requires more 
components. With the dominating adiabatic potential, the width of a given three-body resonance 
can be estimated as
\begin{eqnarray}
T=\exp \left\{ -2\int_{\rho_i}^{\rho_0}\left[ \frac{2m}{\hbar^2}(V_{ad}(\rho)-E_R)\right]^{1/2}d\rho \right\} ~ \ ,
\label{wkb}
\end{eqnarray}
where $E_R$ is the energy of the resonance. $\rho_i$  and $\rho_o$ in Eq. (\ref{wkb}) are the 
inner and outer classical turning points defining endpoints of the pass through the barrier. If 
crossings of various adiabatic potentials occur along the way through the barrier, the system 
is supposed to follow a path corresponding to the smoothest adiabatic potential.

The Hyperspherical Adiabatic Expansion (HAE) method has been extensively used to study different possible situations 
in three-body decay resulting from various combinations of short and long-range interactions and different 
binding-energy situations of all two-body subsystems~\cite{garrido05,garrido05a,garrido06}. In particular, 
it has been applied for the description of the Borromean nucleus 
$^{17}{\rm Ne} (^{15}{\rm O}+{\rm p}+{\rm p})$~\cite{garrido05,garrido04}.  In this application, a 
parameterized proton-proton interaction was used~\cite{garrido97} and the proton-core interaction was described 
by an $\ell$-dependent potential with central, spin-spin and spin-orbit radial potentials. 

The HAE approach allows to distinguish between sequential and direct three-body decays. Large 
separation between all pairs of particles in the final state 
can be achieved either by a uniform separation of all particles or by first moving one particle 
to infinity while the others remain at essentially the same distance from each other and second 
by moving the two close-lying particles apart. The latter scenario defines the sequential process 
where the second step of a two-step process starts after the first particle is at a distance larger 
than the initial size of the three-body system. In $^{17}$Ne, both the 
ground state (1/2$^-$) and the first excited state (3/2$^-$) have a lower energy than the proton-unstable 
ground state of $^{16}$F. Hence, the standard sequential decay process through the intermediate 
resonance(s) in $[A-1]$ nucleus is not possible. However, nothing prevents the decay from 
proceeding via the energetically favourable path described by the lowest adiabatic potential 
until at some point the energy conservation dictates that also this two-body structure must 
be broken. This mechanism which resembles the virtual sequential decay process discussed 
above (see Sect. \ref{sub_seq}) leads to a  significant reduction of the 2p partial half-life 
of  the 3/2$^-$ state in $^{17}$Ne~\cite{garrido05,garrido04}.

Principal assumptions of the HAE approach are (i) the adiabatic  motion in the three-body 
system, and (ii) the cluster ansatz for the many-body wave function of the total system. The latter 
assumption will be discussed in Sect. \ref{grig_theor} in connection with another three-body cluster model, 
the so-called Systematic $\ell^2$ Model (${\rm S}\ell^2{\rm M}$)~\cite{grigorenko03d,grigorenko03c}.
The adiabaticity assumption is better satisfied for certain collective states and for a spontaneous fission 
process. However, this hypothesis becomes questionable if several adiabatic potentials 
are crossing as found in charged-particles/fragments emission~\cite{garrido05,garrido04}.

\paragraph{ The three-body cluster model: the Systematic $\ell^2$-Model  \label{grig_theor}}

The ${\rm S}\ell^2{\rm M}$ of  Grigorenko et al.~\cite{grigorenko03d,grigorenko03c} is a flexible tool to 
simulate various consequences of 
three-body dynamics in 2p emission decay. This model  is a variant of a three-cluster model 
with outgoing flux. Like in HAE approach, the 
${\rm S}\ell^2{\rm M}$ uses hyperspherical harmonics to treat the decay of a 
three-body system. However, unlike the HAE method, the ${\rm S}\ell^2{\rm M}$ includes essentially 
only the direct three-body decay. In general, the three-cluster models are useful for understanding the structure 
of certain states in light nuclei, such as $^6$He, $^6$Li, $^6$Be~\cite{zhukov93}, 
$^8$Li, $^8$B~\cite{grigorenko98,grigorenko99}, and $^{11}$Li~\cite{zhukov93}. 

The basic assumption in ${\rm S}\ell^2{\rm M}$ is that the intrinsic constituent particle 
degrees of freedom can be implicitly treated  through the judicious choice of both the active 
few-body degrees of freedom and the appropriate effective two-body interactions. The cluster 
ansatz is physically more relevant for states close to the cluster-decay 
threshold~\cite{ikeda68,okolowicz03,luo02,michel05,okolowicz06}. Hence, the three-cluster models are perhaps 
better suited for halo systems. On the other hand,  couplings to the
decay channels involving intrinsic states of a daughter nucleus with excitation energies up to $\sim 12$~MeV~\cite{luo07} 
are strongly present for near-threshold states. 
Therefore, even in halo states the wave function is a superposition of several different cluster 
configurations.

In heavier nuclei, the halo wave functions are  too complex to be reduced into a simple 
three-cluster parameterization. Nevertheless, the  ${\rm S}\ell^2{\rm M}$ 
provides in those nuclei an order of magnitude estimate of the relation between the three-body decay energy, 
the effective interactions in the proton-proton and proton-core subsystems, 
and the probabilities of the 
$\ell^2$ configuration of the 2p emitter. Such estimates are very helpful in phenomenological 
applications. Moreover, by solving decay kinematics in hyperspherical variables, the  
${\rm S}\ell^2{\rm M}$ allows to study angular correlations and energy/momentum correlations between decay 
products which presently cannot be investigated by any other approach.

In ${\rm S}\ell^2{\rm M}$, one begins by calculating the internal wave function of the system 
$\Psi_{\rm int}$ at a discrete value of the energy $E_{\rm int}$. This wave function is a 
solution of the Schr{\"o}dinger equation: $$({\hat H}-E_{\rm int})\Psi_{\rm int}=0$$ with 
zero boundary condition at some finite value of the hyperradius $\rho_{int}$. The time-independent 
part of the wave function for decay particles $\Psi^{(+)}$ with purely outgoing asymptotics and 
an energy $E_{\rm int}$ is found by solving the inhomogeneous equation 
$$({\hat H}-E_{\rm int})\Psi^{(+)}=-i\Gamma/2\Psi_{\rm int}$$ with the source term given by an
internal wave function. This way of modelling  is based on the assumption 
that the decaying state is a narrow 
resonance with a negligible coupling to one- and two-particle continua of decay channels.
This assertion is questionable for couplings to one-particle decay channels, either closed 
or opened, which induce the dynamical effects in the 2p decay. 

The 2p separation energy in ${\rm S}\ell^2{\rm M}$ is adjusted using a  weak short-range 
three-body force. 
It is assumed that this force does not modify the barrier penetration process. 
An essential element of the ${\rm S}\ell^2{\rm M}$ is the proton-proton 
final-state  interaction which is tuned to the proton-proton phase shift.   
In the proton-core channel, adjustable potentials of the Woods-Saxon  form are employed. The 
internal structure of the parent nucleus $[A]$ is tuned by changing parameters of the proton-core 
interaction potential for different $\ell$ values. In this way, one generates wave functions 
with different populations of $\ell^2$-configurations. Clearly, 
this treatment excludes genuine many-body effects, such as the configuration interaction and 
configuration mixing, or the Pauli blocking which is approximated  in ${\rm S}\ell^2{\rm M}$ by repulsive 
cores for occupied orbitals. 

Both  ${\rm S}\ell^2{\rm M}$ and HAE approaches are 
precisely tuned to describe the three-body structures and decays. This gives them a certain conceptual 
advantage over SM-like approaches. On the other hand, an internal structure of the decaying state is 
oversimplified, the treatment of Pauli's exclusion principle is incomplete, and the information about 
the complexity of many-body resonance  wave function is lost in adjusted interaction potentials.  

Certainly, some of deficiencies of three-body cluster models can be hidden by a judicious  choice 
of phenomenological potentials. However,  the calculated results can hardly be more reliable 
than the basic assumptions. Due to an unrealistic structure of the decaying 2p pair in 
${\rm S}\ell^2{\rm M}$, the relative weight of various $\ell^2$-configurations 
obtained by fitting the experimental 2p half-life cannot be used to extract meaningful 
information about proton-proton correlations inside a nucleus. It is then astonishing 
that the ${\rm S}\ell^2{\rm M}$ can give an order of magnitude estimate of the experimental data.

\section{Comparison between experiment and model predictions}

Results of configuration-interaction approaches  (SMEC and extended R-matrix approach) can be compared 
to evaluate the role of nuclear dynamics and continuum couplings in the 2p decay. Unfortunately, 
no meaningful comparison is possible between SM-like approaches and  
three-body cluster models (HAE approach or ${\rm S}\ell^2{\rm M}$).

In this section, we shall discuss results of SMEC and extended R-matrix approach for 
$^{45}$Fe, $^{48}$Ni and $^{54}$Zn assuming the diproton scenario of 2p radioactivity. 
We shall also evaluate in SMEC an importance of the virtual sequential  process in 
these decays. The experimental evidence for the 2p 
radioactivity in $^{45}$Fe, $^{48}$Ni and $^{54}$Zn has been discussed in 
Sects. 6.2, 6.4 and 6.3, respectively (see also Table 1).

\subsection{The decay of $^{45}${\rm Fe} \label{sect_dipr_Fe}}

In  SMEC studies of $^{45}$Fe, internal many-body states in ${\cal Q}$ are obtained by solving the standard 
SM with the IOKIN Hamiltonian ~\cite{nummela01}. 
Configurations with excitations from the $sd$ shell to the $fp$ shell are excluded. The residual couplings between 
${\cal Q}$ subspace and the embedding continuum is given by the Wigner-Bartlett interaction:  
$V^{(res)}={\bar V}_{0}[\alpha + \beta P^{\sigma}]\delta(\bf{r_{2}}- \bf{r_{1}})$, 
with $\beta=0.27$ ($\alpha+\beta=1$) and ${\bar V}_0=-900$ MeV$\cdot$fm$^{3}$. $P^{\sigma}$ is the spin-exchange operator.

The calculated diproton half-life  is $13.3_{-4.9}^{+8.2}$ms~\cite{rotureau06}. The error bar is associated with 
an experimental uncertainty on the decay energy 
$Q_{2p}$. External mixing of SM wave functions due to the ${\cal Q}$-${\cal T}$ couplings 
gives a negligible correction. More important are the ${\cal Q}$-${\cal P}$ couplings. The external
mixing generated by these couplings reduces the diproton half-lives  by about 
10\% for $Q_{1p}$ in the interval $-0.1{\rm MeV}<Q_{1p}<0.1{\rm MeV}$. 
This reduction depends weakly both on the magnitude of $Q_{1p}$ and its sign. The experimental partial 
half-life for 2p radioactivity is by factor $\sim 4$ shorter (cf Table 1).  

The extended R-matrix model  yields the diproton half-life of $46_{-16}^{+25}$ms~\cite{brown03}. The nuclear 
structure enters in this model through $^2$He spectroscopic factors which are calculated in $fp$ shell using 
the GPFX1 Hamiltonian~\cite{honma02,honma04}. The calculation using FPD6 interaction~\cite{richter91} yields 
a similar value of the half-life. This half-life is by a factor $\sim 4$ slower than the SMEC prediction. 

The sequential 2p decay is always a part of  the 2p decay, independently of the sign and magnituude of 
$Q_{1p}$. One should remind that the formal separation between the diproton decay and the virtual sequential  
emission cannot be justified if these two modes yield comparable partial decay widths, i.e. 
$\Gamma_{2p}^{(dir)}\simeq\Gamma_{2p}^{(seq )}$.

Let us consider the sequential emission of $^{45}$Fe through the $2^-$ and $3^-$ continua associated with 
the ground state ($J^{\pi}=2^-$) and the first excited state ($J^{\pi}=3^-$) of $^{44}\rm{Mn}$, respectively. 
For $Q_{1p}<Q_{2p}$, both $J^{\pi}=2_1^-$ and $J^{\pi}=3_1^-$ states are resonances which decay by 1p emission. 
Details of calculations can be found in Ref.~\cite{rotureau06}. 

\begin{figure}[hht]
\begin{center}
\includegraphics[height=14cm,width=10cm,angle=-90]{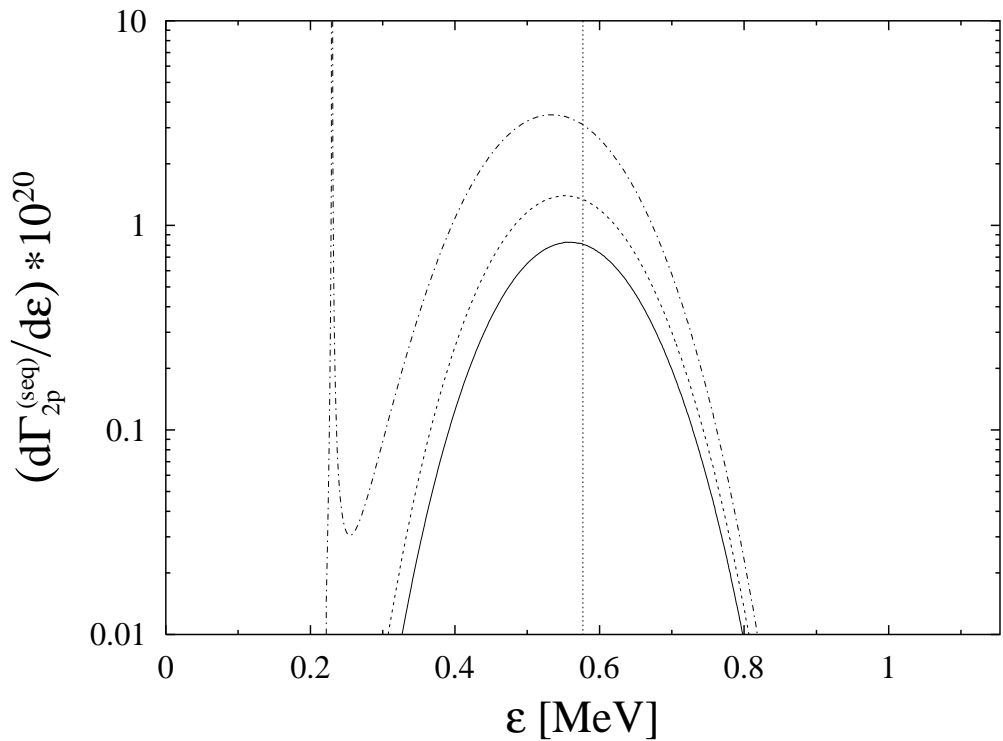}
\end{center}
\caption{The energy distribution of the first emitted proton in a sequential decay of   
      $J^{\pi}=3/2^+$ ground state of $^{45}\rm{Fe}$.  SMEC calculations, including effect of an external 
      ${\cal Q}$-${\cal P}$ mixing,  have been performed for different 
      energies of $J^{\pi}=2^-$ ground state of $^{44}{\rm Mn}$: $Q_{1p} = -0.1$ MeV (solid line), 
      $0.05$ MeV (dashed line) and $0.23$ MeV (dashed-dotted line). 
       The vertical line denotes $Q_{1p}=Q_{2p}/2$. From Ref. \cite{rotureau06}.}
\label{seq_dep_eps_fig}
\end{figure}

The sequential emission half-life depends strongly both on the position of 1p emission 
threshold ($Q_{1p}$) and on the total decay energy ($Q_{2p}$). Assuming  that $^{45}$Fe 
is stable with respect to 1p emission ($Q_{1p}=-0.1$MeV), the virtual sequential 2p half-life 
is $165_{-58}^{+94}$~ms, i.e. about ~10 times slower than the diproton decay. This estimate
includes an effect of external  mixing of $J^{\pi}=3/2^+$ SM wave functions, which reduces the virtual 
sequential decay half-life by about 30\%. For $Q_{1p}=+0.05$MeV, even though the standard sequential 
process through an intermediate resonance is possible, its 
contribution is totally screened by the Coulomb barrier. The dominant decay mechanism remains the virtual 
sequential process through  the  'ghost' of ground state of $^{44}$Mn, far away 
from the resonance region $Q_{1p}\pm\Gamma_{2_1^-}/2$ (see Fig. \ref{seq_dep_eps_fig}). The 
half-life $T_{1/2}=110_{-29}^{+61}$~ms is about a factor of 8 slower than the diproton decay.  
Couplings to the decay channels $(3^-,l_j)^{3/2^+}$ associated with an excited state of $^{44}$Mn 
are relatively important and further reduce the half-life by about 30\% for $Q_{1p}=0.05$ MeV and 
about 38\% for $Q_{1p}=-0.1$ MeV. One cannot exclude that also decay channels connected with higher 
lying states of $^{44}$Mn contribute significantly to the reduction of the 2p partial 
half-life~\cite{luo07}.

\begin{figure}[hht]
\begin{center}
\includegraphics[height=14cm,width=10cm,angle=-90]{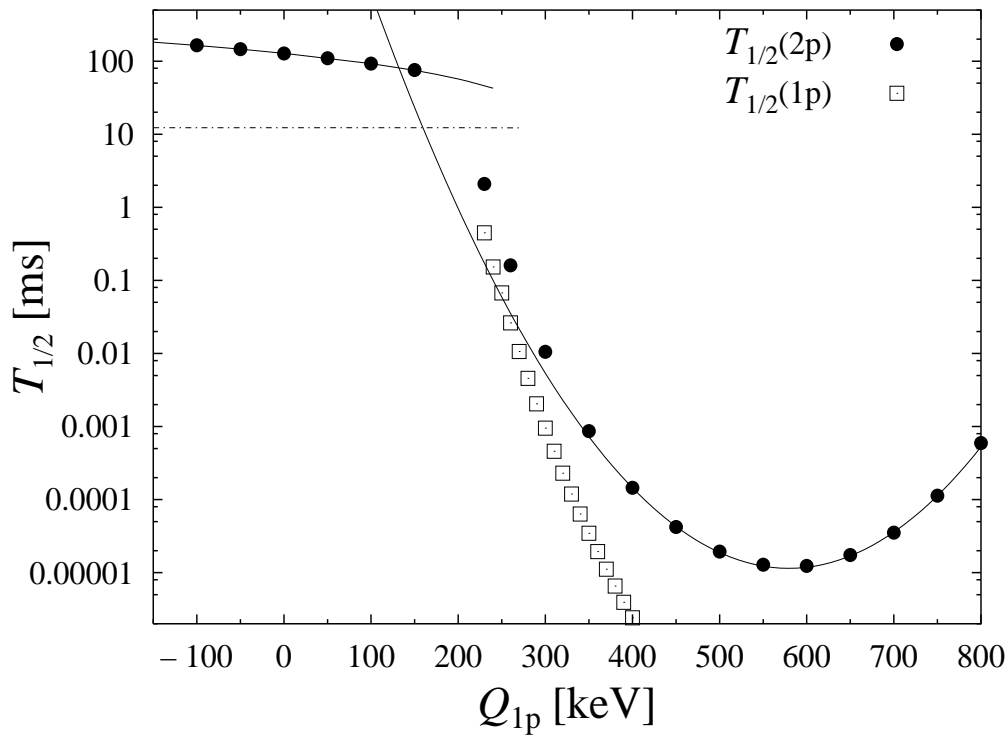}
\end{center}
\caption{The half-lifes for sequential 2p decay (full points) and 1p decay (squares) of  $^{45}\rm{Fe}$. 
      SMEC calculations include effects of an external 
      ${\cal Q}$-${\cal P}$ mixing. The dashed-dotted line shows the half-life for a diproton decay. From Ref. \cite{rotureau06}.}
\label{seq_dep_Q1_fig}
\end{figure}

An energy of the first emitted proton is limited by $Q_{2p}$. The most probable energy of first proton 
in the  virtual sequential decay~\cite{rotureau06}
($\varepsilon_{1p}^{(ghost)}\leq 0.55$ MeV) slightly  decreases with increasing $Q_{1p}$ (see Fig. \ref{seq_dep_eps_fig}). 
A small asymmetry seen in this distribution is a result of different  barriers for proton emissions in $^{45}$Fe 
and in $^{44}$Mn. The full width at half maximum of a ghost is $\Gamma^{(ghost)}=0.197$ MeV at 
$Q_{1p}=-0.1$ MeV.  $\Gamma^{(ghost)}$ increases with $Q_{1p}$
in the range of $Q_{1p}$ values where the virtual 2p decay dominates. Experimental observation of symmetric 
energy distribution of individual protons~\cite{miernik07b} indicates that either $Q_{1p}$ in  $^{45}$Fe is 
less than $\sim -0.25$ MeV, or the diproton decay mechanism dominates. The latter scenario is favoured for 
typical effective SM interactions and $Q$-values. One should stress that these conclusions have been obtained 
assuming a coherent contribution of diproton and virtual sequential decay mechanisms.

For $Q_{1p}<0.2$ MeV,  the sequential 2p decay is predominantly related to the strength of a ghost in the energy interval 
$\varepsilon_{1p}^{(ghost)} - \Gamma^{(ghost)}/2 < \varepsilon < \varepsilon_{1p}^{(ghost)} + \Gamma^{(ghost)}/2$ 
and not to the strength of 1p resonance at $Q_{1p}$. In this regime, the sequential 2p decay 
width does not reduce to the product of the width for the first step 
($\Gamma_{1p}$) and the branching ratio for the second step, as one would obtain semi-classically. 
Above $Q_{1p}\simeq0.23$ MeV, the sequential decay is dominated by the standard two-step process 
via the ground state of $^{44}\rm{Mn}$. 

The $Q_{1p}$-dependence of the sequential 2p decay half-life 
is shown as full points in Fig. \ref{seq_dep_Q1_fig}. Squares in Fig. \ref{seq_dep_Q1_fig}
represent the 1p decay half-life and the diproton half-life is depicted by the dashed-dotted line. 
The sequential 2p emission and the diproton emission, considered as independent, non-interfering decay paths, 
yield comparable half-lives for $0.15~{\rm MeV}<Q_{1p}<0.2$ MeV. 
In this interval, the three-body 2p decay cannot be reduced to any sequence of two-body decays. 

An interval of $Q_{1p}$ values for which the half-lifes of virtual sequential and diproton decays are expected 
to become similar lies close to the range of predicted 
$Q_{1p}$-values~\cite{brown91,ormand96,cole96}.  Hence, 
both processes could interfere strongly and, in spite of symmetric energy distribution of individual protons~\cite{miernik07b}, 
the measurement of $Q_{1p}$ in  $^{45}$Fe is mandatory before drawing definitive conclusions about the 2p emission mechanism 
in this nucleus.

The sequential 2p decay has two distinct dependences of the decay half-life on $Q_{1p}$ (see  Fig.  \ref{seq_dep_Q1_fig}). 
For small positive values of $Q_{1p}$ ($0<Q_{1p}\leq 0.2$ MeV) as well as for closed 1p decay channels, the sequential 
2p decay half-life changes linearly with $Q_{1p}$. This is the regime of  virtual sequential 2p decay, 
where $\Gamma_{2p}^{(seq)}\gg \Gamma_{1p}$ and the 2p decay goes via the 'ghost' of the ground state. 
For $Q_{1p}> 0.23$ MeV, one enters in the regime $\Gamma_{1p}\gg \Gamma_{2p}^{(seq)}$, 
where the role of $2_1^-$ resonance in $^{44}$Mn is dominant. In this regime of standard sequential decay, the 2p partial half-life:
$$T_{1/2}(Q_{1p})\sim\exp -(Q_{1p}-Q_{1p}^{(0)})^2 $$ 
 has a minimum at $Q_{1p}^{(0)}\sim Q_{2p}/2$.  

\subsection{The decay of $^{48}${\rm Ni} \label{sect_dipr_Ni}}

In Ref.~\cite{rotureau06}, several SM Hamiltonians including IOKIN~\cite{nummela01}, 
KB3~\cite{poves81} or GXPF1~\cite{honma02,honma04} were used to 
evaluate the sensitivity of  SMEC results to the choice of an effective interaction.
For  the IOKIN Hamiltonian, the diproton half-life is $6.2_{-2.45}^{+4.1}$ms. 
The error bars are related with an experimental uncertainty of the decay energy $Q_{2p}$. 
External mixing of SM 
wave functions generated by the ${\cal Q}$-${\cal T}$ and ${\cal Q}$-${\cal P}$ 
couplings are unimportant, reducing the diproton half-life by $\sim 2$\% in the interval 
$-0.1{\rm MeV}<Q_{1p}<0.1{\rm MeV}$. Somewhat stronger is the dependence of 2p partial half-lifes 
on the SM interaction. For GXFP1 and KB3 interactions, one finds $7.4_{-2.9}^{+4.9}$ms and 
$6.9_{-2.7}^{+4.5}$ms, respectively. For all three SM interactions, the SMEC results are consistent 
with the experimental partial half-life. The extended R-matrix model  with GXFP1 interaction yields a 
diproton half-life $16_{-4}^{+10}$ms~\cite{dossat05} which  is about a factor of 2 slower than the SMEC result. 

In the description of virtual sequential emission, the 2p emission through the $3/2^-$ and $7/2^-$ continua 
associated with the ground state ($J^{\pi}=3/2^-$) and the first excited state ($J^{\pi}=7/2^-$) of $^{47}$Co 
were considered. SMEC calculations were performed with the  IOKIN Hamiltonian. Principal 
uncertainties of the SMEC calculations  are (i) an unknown position of 1p emission 
threshold in $^{48}$Ni, 
and (ii) an insufficient precision of the total decay energy.  If $Q_{1p}=-0.1$MeV, then the virtual 
sequential 2p half-life is $25.4_{-8.9}^{+14.3}$ms, i.e. about ~4 times slower than the 
diproton decay. This estimate includes an external mixing of $3/2^+$ SM wave functions generated 
by the ${\cal Q}$-${\cal P}$ couplings. These couplings reduce the virtual sequential decay half-life 
by about 15\%. For $Q_{1p}=+0.05$MeV, the standard sequential decay through an intermediate resonance 
in $^{47}$Co is totally screened by the Coulomb barrier. Hence, the dominant decay mechanism remains 
the virtual sequential process through the 'ghost' of $^{47}$Co ground state. 
The half-life in this case 
($T_{1/2}=16.5_{-5.7}^{+9.2}$ms) is only $\sim 2.5$ slower than the diproton decay.  
Couplings to the decay channels $(7/2^-,l_j)^{0^+}$ associated with the excited state of $^{47}$Co are 
relatively unimportant, reducing the half-life by $\sim 6\%$ - $9\%$.

\subsection{The decay of $^{54}${\rm Zn} \label{sect_dipr_Zn}}

The diproton half-life  of $^{54}$Zn in SMEC with GXFP1 Hamiltonian~\cite{honma02,honma04}  is 
$13.8_{-5.1}^{+8.4}$ms. The error bar is due to the experimental uncertainty on the decay energy $Q_{2p}$~\cite{rotureau06}. 
With KB3 Hamiltonian, the diproton half-life  is somewhat longer and equals $17_{-6.3}^{+10.3}$ms ~\cite{poves81}. 
The experimental partial half-life for 2p radioactivity is about a factor of 4 faster. As in  $^{45}$Fe and $^{48}$Ni, 
a significant contribution from virtual sequential 2p decay cannot be excluded but, again, the quantitative estimate 
of this contribution cannot be made without knowing $Q_{1p}$ in $^{54}$Zn. 

The extended R-matrix model  with GXFP1 interaction yields the diproton half-life $10_{-4}^{+7}$ms~\cite{blank05zn54} 
which is about 30\% shorter than the SMEC prediction. 

\subsection{Discussion  \label{disc_concl}}

The coupling of a nuclear system to the environment of its decay channels may change its properties 
in a non-perturbative way~\cite{okolowicz03}. The magnitude of the system - environment 
coupling depends 
explicitly on the location of various emission thresholds and on the structure of 
$S$-matrix polers. The description of resonances and their decays are mainly sensitive to the two 
ingredients: the effective Hamiltonian ${\cal H}_{QQ}$ (see Sect. \ref{SMEC_1p}) and the 
unitarity of the $S$-matrix which causes a non-trivial energy dependence of the coupling 
matrix elements between resonance states and continuum.  ${\cal H}_{QQ}$ includes basic 
structural information contained in $H_{QQ}$ (the closed quantum system Hamiltonian) and the 
coupling matrix elements between discrete and continuous states which depend on the structure of the decay channels. 

SMEC studies of 2p decays  in $^{45}$Fe, $^{48}$Ni and $^{54}$Zn have shown that the virtual 
sequential 2p emission cannot be separated easily from the diproton decay, even though they 
describe two different limits. The first one corresponds to two uncorrelated proton emissions separated in time, 
whereas the latter one represents an instantaneous  
emission of strongly correlated two protons. The most probable energies of these protons are nearly the same in both scenarios. 

If half-lives for these two pathways are strongly different, then  neglecting 
their mutual interference and the omission of spatio-temporal correlations in the virtual 
sequential decay can be probably justified. Otherwise, the 2p emission described by an effective 
Hamiltonian (\ref{H_eff_T2})  has to be considered as 
a genuine three-body decay. Unfortunately, the 1p emission thresholds in $^{45}$Fe, 
$^{48}$Ni and $^{54}$Zn are unknown so that one cannot rule out the possibility of a 
strong interference between (virtual) sequential and diproton decay modes. SMEC calculations 
in $^{45}$Fe with $Q_{1p}$ values which are consistent with available theoretical estimates  
~\cite{brown91,ormand96,cole96} show that the virtual 2p sequential half-life is by a factor of about 10 
longer than the diproton half-life. This would favour an interpretation of the observed 2p 
emission in this nucleus as predominantly the diproton decay. Such a conclusion agrees also with the observation 
of a symmetric energy distribution of individual protons in  the decay of $^{45}$Fe~\cite{miernik07b}.

For the sake of argument, let us suppose that the virtual sequential emission mechanism yields 
much longer half-lives in all three nuclei: $^{45}$Fe, $^{48}$Ni and $^{54}$Zn. It is then 
interesting to notice that the diproton decay mechanism in SMEC reproduces quite well the 
experimental partial 2p decay half-life in a doubly magic nucleus ($^{48}$Ni) and 
fails by a factor of about 4 in both open-shell nuclei ($^{45}$Fe and $^{54}$Zn). If this 
tendency is not coincidental, or follows from a poor knowledge of 
$Q$-values and an insufficient experimental statistics, then one might be tempted to associate it with
pairing correlations. As suggested in Ref.~\cite{blank05zn54}, the change of
pairing correlations induced by the configuration mixing beyond the 1p0f shell may well 
be responsible for this enhancement. A similar effect has been seen a long time ago in (p,t) 
transfer reactions~\cite{decowski78}. The absence of this component of pairing correlations is 
more important in open-shell nuclei than in a closed-shell nucleus $^{48}$Ni. Hence, one may 
expect that the fp shell SMEC calculations are better suited to describe the observed 2p partial 
half-life in $^{48}$Ni than in $^{45}$Fe and $^{54}$Zn. 

One should mention that different effective interactions in the $fp$ and $sdfp$ shells give similar 
results to within $\sim 10-20$\%. External mixing, generated by the ${\cal Q}$-${\cal P}$ 
coupling, modifies the diproton decay width by less than 10\% in $^{45}$Fe and $\sim 2$\% 
in $^{48}$Ni. Again, the SMEC predictions for the closed-shell nucleus $^{48}$Ni seem to be more reliable since 
the direct $Q_{1p}$-dependence of the diproton half-life is weaker in this case. 

The R-matrix model ~\cite{brown03} calculations show a gradual decrease of the 2p partial half-life 
going from $^{45}$Fe to $^{54}$Zn across the shell closure at Z=28, N=20. Consequently, the 
diproton half-life in $^{45}$Fe is by a factor of $\sim 5$ longer than in $^{54}$Zn. On the contrary, 
both in the SMEC results and in the data, the 2p partial half-lives of $^{45}$Fe and $^{54}$Zn are very close. 

\begin{figure}[hht]
\begin{center}
\includegraphics[height=12cm,width=17cm,angle=00]{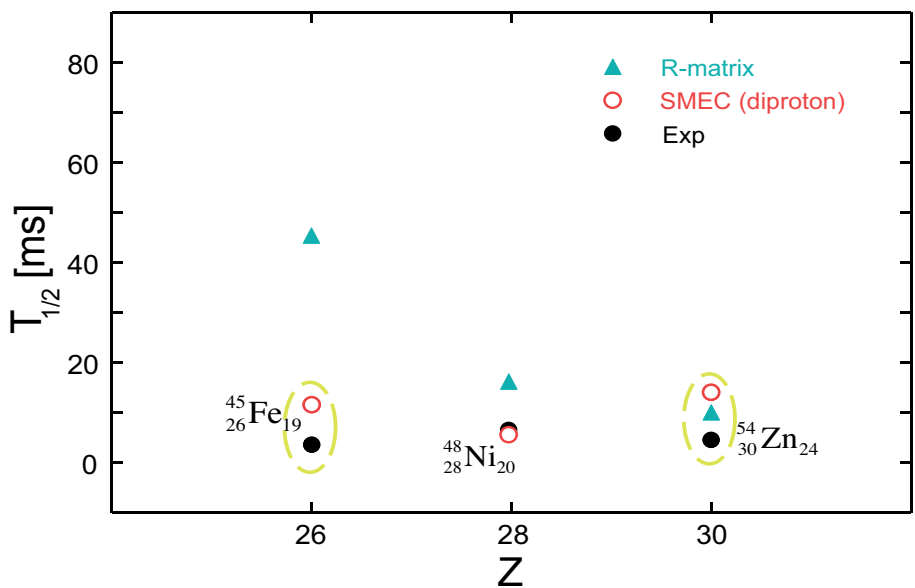}
\end{center}
\vspace{-2cm}
\caption{The partial half-lives for the diproton decay from the ground state of  $^{45}_{26}\rm{Fe}$, 
      $^{48}_{28}\rm{Ni}$, and $^{54}_{30}\rm{Zn}$, as calculated in the generalized R-matrix theory 
     ~\cite{brown03} (full triangles) 
      and in the SMEC~\cite{rotureau06}(open circles). The experimental values are shown with black points. For
      reasons of clarity, the uncertainty of theoretical results due to the uncertainty 
      of the $Q_{2p}$ values and experimental error bars are not shown.}
\label{systematic}
\end{figure}

Fig. \ref{systematic} compares the experimental partial half-lives for the diproton decay of  
$^{45}_{26}\rm{Fe}$, $^{48}_{28}\rm{Ni}$, and $^{54}_{30}\rm{Zn}$ with the calculated ones in the 
SMEC and in the extended R-matrix model. One may notice two characteristic features from this 
comparison. Firstly,  SMEC and R-matrix models predict a different Z-dependence of calculated 
2p partial half-lives around the doubly-magic nucleus $^{48}$Ni. Secondly, SMEC yields an opposite
tendency of $^{45}$Fe and $^{54}$Zn half-lives with respect to the half-life of $^{48}$Ni than 
seen experimentally. Concerning the first point, one may tentatively assume that this feature is related 
to the continuum coupling induced dynamical effects in the 2p decay which are missing in the 
R-matrix description. Certainly, one does not expect that this different systematic tendency 
of calculated half-lives  in the SMEC approach and in the R-matrix model could be explained by the 
absence of configuration mixings beyond the 1p0f shell, since both in SMEC and in R-matrix calculations 
these configurations are not included. Concerning the second point, it may be related to the small number 
of channels in SMEC calculations of the ${\cal Q}$-${\cal P}$-coupling induced an external configuration mixing. In 
Ref.~\cite{rotureau06}, the external mixing is a result of coupling between the ground state of nucleus 
$[A]$ and the two channels corresponding to the ground state and the first excited state of a nucleus 
$[A-1]$. This restriction on the number of channels is justified in closed-shell nuclei as 
in this case the ground-state to ground-state coupling of the $[A]$ and $[A-1]$ nuclei exhausts up 
to about 90\% of the total coupling strength. For open-shell nuclei, the strength of continuum coupling is largely 
spread over many channels, both open and closed. Indeed, the SMEC studies of continuum 
coupling effects in oxygen and fluorine isotopes~\cite{luo07} have shown
that a contribution of the ground-state to ground-state coupling to the total 
continuum coupling strength may vary between a few $\%$ and $\sim 30\%$ in open-shell nuclei. 
The size of a model space involved in the calculation of the 2p decay in $^{45}_{26}\rm{Fe}$, 
$^{48}_{28}\rm{Ni}$,  and $^{54}_{30}\rm{Zn}$ makes it impossible to verify this observation numerically, 
but if true it may lead to a strong enhancement of the contribution from the virtual 
sequential 2p decay path and, henceforth, to a decrease of the  2p partial half-life. One should remind, 
that the interference between the virtual sequential and diproton emission paths cannot be probably neglected 
if both processes become equally important.

As said before, SMEC predictions are very sensitive to $Q$ values which for the  
nuclei studied are either not known experimentally with a sufficient precision ($Q_{2p}$ values), 
or are unknown ($Q_{1p}$ values). The virtual sequential decay half-lives 
depend explicitly and strongly on the continuum coupling dynamics  between the ${\cal Q}$, 
${\cal P}$ and ${\cal P}$, ${\cal T}$ subspaces which is governed by $Q$ values. A direct 
dependence of the diproton half-lives on ${\cal Q}$-${\cal P}$ virtual couplings is rather 
weak, i.e. dressing of the diproton decay mode  by virtual 1p excitations does not 
change significantly its properties in $^{45}$Fe, $^{48}$Ni and $^{54}$Zn~\footnote{This is in 
contrast to the diproton decay from the second  excited $1^-$ state 
of $^{18}$Ne~\cite{rotureau05}. In this case, the diproton mode is dressed by virtual excitations 
to the {\em opened} 1p emission channels leading to the particle stable $5/2_1^+$, $1/2_1^+$ final states in $^{17}$F. 
In general, couplings to open 1p emission channels strongly modify 
properties of the diproton mode.}. 
On the other hand, this idealized decay mode represents a sensible approximation of the 2p 
decay process described by (\ref{H_eff_T}) only if the virtual sequential decay path is 
strongly suppressed. This in turn depends directly on  the dynamics of the ${\cal Q}$, ${\cal P}$ 
and ${\cal P}$, ${\cal T}$ virtual couplings. Hence, further progress in describing spontaneous 
2p radioactivity and understanding the role of both pairing correlations and virtual excitations 
to continuum states in this process depends on a precise experimental determination of the 
$Q_{1p}$ and $Q_{2p}$ values.

\section{Conclusions and outlook}

The study of 2p emission opens a new window to investigate nucleon-nucleon correlations 
and the structure of atomic nuclei. Although these studies, at least partially, could have been performed 
already with $\beta$-delayed 2p emitters, the field received a strong boost with the observation of 
ground-state 2p radioactivity. This discovery triggered new 
experimental developments, such as time-projection chambers, and new theoretical activity related 
mainly to the unified description  of structure and reactions with two-particle continua and the 
development of various methods to deal with three-body asymptotic states of charged particles/fragments.

Both experiment and theory have reached a certain level of sophistication, but significant improvements 
are still needed to develop 2p radioactivity into a powerful tool for nuclear structure studies. On the 
theoretical side, the lack of three-body asymptotic channels hampers the calculation of various correlation 
functions in the SMEC framework. Description of  direct  2p emission with three-body asymptotics in SMEC 
requires inclusion of  finite-range interactions. 
At the moment, the angular correlations between decay products can be studied 
solely in the three-body cluster models. The simplified description of nuclear structure 
in cluster models does not allow to draw deeper conclusions about 2p 
in-medium correlations from these angular correlations.

As one deals with three particles in the final state, the 2p emission can occur in different ways, i.e. the 
asymptotic channel cannot be defined unambiguously. Consequently and in 
contrast to one-proton or $\alpha$ radioactivity, the 2p radioactivity does not have a unique experimental 
signature and its interpretation is always depending on the theoretical interpretation. A complete 
theoretical description of 2p emission has to take into account all different decay paths and include 
interferences between them. It is not obvious whether different 2p decay paths, such as the diproton 
path and the virtual sequential 2p decay path which both yield similar proton energy distributions, 
can be distinguished by, e.g., the proton-proton and/or proton-fragment correlations. Further studies 
along this line are necessary for a best possible characterization of 2p radioactivity.

On the experimental side, more cases of 2p radioactivity have to be studied and the quality of the data
has to be improved, as the possibilities to compare the experimental data existing today with modern theories 
of 2p radioactivity are hampered by large experimental error bars. To improve the experimental data, increased 
statistics is only one aspect. In the case of $^{45}$Fe, the major contribution to the uncertainty of the decay 
energy comes from the energy calibration. Therefore, new and improved calibration tools have to be developed as 
well. A precise determination of $Q_{1p}$ and $Q_{2p}$ would help to check different features of SMEC calculations, 
such as the choice of effective interactions and valence space, or the absence of explicit three-body asymptotics. 

All experiments performed up to now seem to indicate that the sequential decay prevails, as long as intermediate 
states for 1p emission are energetically accessible. The only exceptions seem to be the decays of excited states 
in $^{17}$Ne and in $^{94}$Ag$^m$, where despite many intermediate states
a strong proton-proton correlation was observed. A halo-like structure and/or dynamical effects associated with 
the tunnelling through the anisotropic Coulomb barrier could be at the origin of the observed correlation. 
Higher statistics data and advanced theoretical studies are needed to confirm or reject this interpretation. 

Another application of the 2p studies are direct resonant 2p radiative capture processes~\cite{grigorenko05a,grigorenko06}. 
It was demonstrated that the astrophysical 2p capture rates can be enhanced by as much as several orders of magnitude 
in certain ranges of excitation energies. Similar tendency was found for the  size of a nonresonant  E1 contribution 
to the 2p capture rates~\cite{grigorenko06}. These predictions may have important consequences for the rp-process 
path changing earlier predictions concerning rp-process waiting points~\cite{gorres95}. 

The advent of new radioactive beam machines like the RIBF facility in RIKEN, Japan or the FAIR facility
in GSI, Germany should give access to new ground-state 2p emitters such as $^{59}$Ge, $^{63}$Se, 
or $^{67}$Kr  and allow one to achieve much higher rates for the known ones.
These possibilities will most likely significantly increase the body of experimental data and may establish 
2p radioactivity as a powerful tool of nuclear structure studies. 

These new facilities may allow also to search for a similar phenomenon which is one- and two-neutron radioactivity. 
From an experimental point of view, it seems to be rather difficult to find good candidates for these new decay modes. 
Due to the absense of the Coulomb barrier for neutrons, only the centrifugal barrier will prevent these neutrons 
from escaping "immediately" from the nucleus. To create a sufficiently high barrier, high-j orbitals are needed which 
form the nuclear ground state only in medium-mass nuclei. However, 
these medium-mass neutron drip-line nuclei are rather 
far away from the valley of stability and extremely difficult to produce. Possibly, high-j isomers could be candidates 
for this kind of radioactivity. 

 From a theoretical point of view, two-neutron emitters would be ideal cases to study.
All theoretical formalisms can be applied to the two-neutron case without major changes. Actually, the two-neutron three-body 
problem can be solved exactly and not approximately as in the three-body case with charged particles. Moreover, 
angular correlations will be directly related to the internal two-neutron correlations without the uninteresting 
correlations due to a possible anisotropy of the Coulomb barrier. 

\section*{Acknowledgments}

The experimental results on two-proton radioactivity were obtained in large collaborations including the CEN Bordeaux-Gradignan, 
GANIL, GSI, University of Warsaw, and other laboratories. B.B. would like to underline the decisive contributions 
of J. Giovinazzo and M. Pf\"utzner to this work.
One of us (M.P.) wishes to thank J. Oko{\l}owicz and J. Rotureau for a fruitful collaboration which led to 
the development of the continuum shell-model for two-proton radioactivity. Stimulating discussions with 
Witek Nazarewicz are kindly acknowledged.

\end{document}